\documentclass[twocolumn,amsmath,amssymb,aps,prb,longbibliography]{revtex4-1}
\usepackage{color,graphicx}
\usepackage{amssymb}
\usepackage{float}
\begin{document}

\title{Soliton Motion in Skyrmion Chains}
\author{N. P. Vizarim$^1$, J. C. Bellizotti Souza$^2$, C. J. O. Reichhardt$^3$, C. Reichhardt$^3$, M. V. Milo\v{s}evi\'c$^{4,5}$ and P. A. Venegas$^2$}
\affiliation{$^1$ POSMAT - Programa de P\'os-Gradua\c{c}\~ao em Ci\^encia e Tecnologia de Materiais, Faculdade de Ci\^encias, Universidade Estadual Paulista - UNESP, Bauru, SP, CP 473, 17033-360, Brazil\\
$^2$ Departamento de F\'isica, Faculdade de Ci\^encias, Unesp-Universidade Estadual Paulista, CP 473, 17033-360 Bauru, SP, Brazil\\
$^3$ Theoretical Division and Center for Nonlinear Studies, Los Alamos National Laboratory, Los Alamos, New Mexico 87545, USA\\
$^4$ NANOlab Center of Excellence, Department of Physics, University of Antwerp, Belgium\\
$^5$ Department of Physics, University of Antwerp, Groenenborgerlaan 171, 2020 Antwerp, Belgium}

\date{\today}

\begin{abstract}
Using a particle-based model we examine the depinning motion of solitons
in skyrmion chains in quasi-one dimensional (1D) and two-dimensional (2D)
systems containing embedded 1D interfaces.
The solitons take the form of a particle or hole in a
commensurate chain of skyrmions.
Under an applied drive, just above a critical depinning threshold the soliton
moves with a skyrmion Hall angle of zero.
For higher drives, the entire chain depins,
and in a 2D system
we observe that both the solitons and chain move at zero skyrmion
Hall angle and then transition to a finite skyrmion Hall angle as the drive
increases.
In a 2D system with a 1D interface that is at an angle to the driving direction,
there can be a reversal of the sign of the skyrmion Hall angle from positive
to negative.
Our results suggest that solitons in skyrmion systems could be used as
information carriers in racetrack geometries that would
avoid the drawbacks of finite skyrmion Hall angles.
The soliton states become mobile at significantly lower drives than the
depinning transition of the skyrmion chains themselves. 
\end{abstract}

\maketitle

\vskip 2pc

\section{Introduction}

Solitons are a well-known concept in physics
for describing a nonlinear wave, also called a solitary wave, 
that emerges with unchanged shape and speed
from a collision with a similar pulse
\cite{scott_soliton_1973}.
After Zabuski and Kruskal, as part of their investigation of plasma waves
\cite{zabusky_interaction_1965}, coined the term 
``soliton'' due to the novel properties of these solitary waves,
many other branches of science including applied
mathematics \cite{manukure_short_2021},
chemistry \cite{heeger_solitons_1988,tolbert_solitons_1992}, and 
biology \cite{ciblis_possibility_1997,zhou_biological_1989,davydov_theory_1973}
proved to be fertile ground for 
soliton physics. While solitons are 
related to several important phenomena in science,
such as thermal and electrical conductivity, one of 
the areas most impacted by the soliton concept was optics.
Soliton studies greatly enhanced
the technology of optical fibers
\cite{hasegawa_transmission_1973,mollenauer_experimental_1980},
photorefractive crystals \cite{duree_observation_1993}, and
optical media \cite{torruellas_observation_1995}. 
Most recent
studies of solitons appear primarily
in optics (light waves) and matter waves.

In magnetism, the nonlinearity of the spin dynamics
produces topologically non-trivial magnetic structures 
\cite{galkina_dynamic_2018},
including a rich variety of solitons, such as: 
one dimensional solitons 
describing the motion of domain walls \cite{slonczewski_dynamics_1972};
two dimensional magnetic vortices
\cite{papanicolaou_dynamics_1991}; magnon drops \cite{kosevich_magnon_1978}; 
and also the two dimensional topological solitons
called skyrmions \cite{muhlbauer_skyrmion_2009}. Although it is 
well known that skyrmions 
are a type of soliton, we show here
that the collective motion of a chain of skyrmions can also produce
a soliton on a different length scale.

Skyrmions are spin textures pointing in all directions
that can be mapped onto a wrapping of a sphere,
forming a topologically stable 
object \cite{nagaosa_topological_2013}.
One of their
most interesting features
is that they can be set into motion by the
application of a spin polarized current 
\cite{jonietz_spin_2010,schulz_emergent_2012,yu_skyrmion_2012,iwasaki_universal_2013,lin_driven_2013}. In
the presence of external driving, skyrmions can 
exhibit a depinning threshold and obey
nonlinear velocity-force relations
\cite{schulz_emergent_2012,iwasaki_universal_2013,lin_driven_2013,liang_current-driven_2015,woo_observation_2016}.
There is great interest in
using skyrmions as information carriers 
for
memory and logic devices\cite{fert_skyrmions_2013,fert_magnetic_2017}
as well as in spintronics 
\cite{wiesendanger_nanoscale_2016} due to their stability and the low currents
required to set them into motion.
Application of skyrmions in actual devices
will require
a better understanding of their behavior, dynamics, and
how to control their 
motion.

A key aspect of skyrmions that distinguishes
them from other overdamped particles is the presence of a 
non-dissipative component
or Magnus term in their equation of motion
\cite{nagaosa_topological_2013,iwasaki_universal_2013,lin_driven_2013,fert_magnetic_2017,iwasaki_current-induced_2013}.
The Magnus term
produces a skyrmion velocity component perpendicular to the
net force on the skyrmion, and it has been proposed
that the Magnus term is responsible for the reduced depinning threshold
exhibited by skyrmions 
\cite{nagaosa_topological_2013,fert_magnetic_2017,iwasaki_universal_2013}.
In the absence of defects in the sample,
an applied external drive combined with the Magnus term
causes the skyrmion to move at an angle
with respect to the driving direction that is called the intrinsic skyrmion Hall angle, 
$\theta_{sk}^{\rm int}$ \cite{iwasaki_universal_2013,yu_real-space_2010,jiang_direct_2017}.
The magnitude of this angle
increases as the ratio of the Magnus term to the damping 
term is increased.
Experimentally, skyrmion Hall angles have been observed that span the range
from a few degrees up to 
very large angles, depending on the system parameters and
the size of the skyrmions 
\cite{jiang_direct_2017,zeissler_diameter-independent_2020,litzius_skyrmion_2017}.

Recently, it was shown that the skyrmion Hall angle
can be manipulated by introducing periodic pinning 
\cite{reichhardt_quantized_2015,vizarim_skyrmion_2020-1,vizarim_skyrmion_2020-2,feilhauer_controlled_2020,vizarim_directional_2021,stosic_pinning_2017}. 
As the
external drive is increased, the skyrmion Hall angle becomes quantized
due to directional 
locking effects very similar to those found
in superconducting vortices \cite{reichhardt_phase_1999} or 
colloidal assemblies
\cite{bohlein_experimental_2012,reichhardt_directional_2004,gopinathan_statistically_2004}
driven over a periodic substrate under a rotating external drive.
In the case of skyrmions,
the direction of the external drive remains 
fixed, but as the magnitude
of the drive increases, the velocity dependence of the skyrmion Hall angle
causes a change in the direction of 
skyrmion motion.
On a locking step, 
the skyrmion Hall angle remains constant as the magnitude of the drive
is varied, while
changes in the skyrmion Hall angles
are associated with dips or cusps in the velocity-force curves.
This behavior provides
a mechanism for controlling the skyrmion motion in a given sample,
since
a fine 
adjustment in the external driving
can produce a large change
in the skyrmion direction of motion.
Several distinct methods have been proposed
for controlling the skyrmion motion,
including periodic pinning \cite{reichhardt_quantized_2015,vizarim_skyrmion_2020,feilhauer_controlled_2020,vizarim_directional_2021,reichhardt_nonequilibrium_2018}, ratchet effects \cite{chen_skyrmion_2019,gobel_skyrmion_2021,ma_reversible_2017,reichhardt_magnus-induced_2015,souza_skyrmion_2021,yamaguchi_control_2020,vizarim_skyrmion_2020-2}, interface guided motion \cite{vizarim_guided_2021,zhang_edge-guided_2022}, strain gradients \cite{yanes_skyrmion_2019}, magnetic field
gradients \cite{zhang_manipulation_2018,casiraghi_individual_2019,everschor_rotating_2012}, temperature gradients \cite{kong_dynamics_2013,wang_rectilinear_2021}, 1D potential wells \cite{purnama_guided_2015,juge_helium_2021}, nanotracks \cite{leliaert_coupling_2018,zhang_magnetic_2015,chen_skyrmion_2017,toscano_suppression_2020}, and skyrmion-vortex systems in a ferromagnet-superconductor heterostructure \cite{menezes_manipulation_2019}.

Commensurability effects are very important in determining
the collective behavior of skyrmions under the influence of periodic pinning.
When the number of skyrmions $N_{sk}$ is an integer multiple
or rational fraction of the number of
substrate minima $N_p$,
we say that the system is commensurate.
Extensive studies of commensurability effects
have shown that they are associated with distinctive behavior in
many systems, including superconducting
vortices\cite{welp_commensurability_2005,reichhardt_commensurate_1998,harada_direct_1996}, 
colloidal particles\cite{mangold_phase_2003}, Wigner crystals\cite{rees_commensurability-dependent_2012},
and vortices in 
Bose-Einstein condensates\cite{pu_structural_2005,tung_observation_2006}. 
Much less work has been done on commensurability effects
in skyrmion systems
\cite{duzgun_commensurate_2020,reichhardt_commensuration_2022}. 
Recently, Reichhardt {\it et al.} \cite{reichhardt_commensuration_2022}
investigated commensuration effects
for skyrmions in periodic pinning
and found that the skyrmion Hall angle is non-monotonic,
dropping to zero at
commensurate states and returning to a finite value
for incommensurate states.

Solitons often appear in commensurate-incommensurate systems
that are near but not in a commensurate state. Here,
there is an ordered lattice 
containing interstitials or vacancies that behave like
kinks or anti-kinks. Under an applied drive,
these kink objects depin
prior to the ordered portions of the lattice,
resulting in a two-step depinning transition in which
interstitials or kinks
move in the driving direction and vacancies or anti-kinks
move in the opposite direction.
The classic example of a system exhibiting this behavior is the
Frenkel-Kontorova model
\cite{Frenkel38,Braun98,Tekic05}.
Kink  dynamics were imaged directly
in colloidal experiments for 2D periodic substrates
just above and below the
1:1 commensurate conditions \cite{Bohlein12},
while numerical studies of the same system
showed a multi-step depinning
process involving kinks and antikinks \cite{Vanossi12}.
Motion of kinks on periodic substrates
has also been studied in 1D cold
atom systems \cite{Benassi11},
1D and 2D frictional systems \cite{Vanossi13,Vanossi20}, and other systems
near commensuration such as superconducting vortices in periodic pinning arrays 
\cite{Reichhardt17}. Kink motion should also be possible
in skyrmion chains near commensurate conditions; however,
due to the non-dissipative Magnus term,
such kinks would have different dynamics than previously
studied kinks.
Most kink systems have overdamped or underdamped dynamics and
the interstitial solitons move in the same direction
as the applied drive.
In a skyrmion system,
the Magnus term
can cause the kink to move at an angle to the driving direction.
Soliton motion in skyrmion chains is of interest since
the solitons themselves, rather than the skyrmions, could serve as
information carriers.
This would be particularly relevant if kinks
move along the driving direction
under drives much lower 
than those that would be needed
to translate individual skyrmions or chains of skyrmions over
long distances.

In this work, we investigate the skyrmion collective behavior just outside
of a commensurate filling for
$N_{sk}=N_{p}+1$ or $N_{sk}=N_{p}-1$.
We use a heterogeneous pinning lattice
containing a line of weaker pinning potentials that serve
as a guide for the skyrmion motion.
We apply an external dc drive to the sample and neglect
thermal effects.
A soliton in a skyrmion chain, formed by an interstitial 
skyrmion for $N_{sk}=N_{p}+1$
or a vacancy for $N_{sk}=N_{p}-1$, can be set into motion by the external drive.
The interstitial skyrmion moves parallel to the applied drive and the vacancy
moves antiparallel to the applied drive, but both have a zero skyrmion Hall angle,
which is of interest for applications.
We show that the line of weak pinning potentials can guide the
soliton motion even to the point of causing the soliton to move
in the direction $-\theta_{sk}^{\rm int}$, opposite to the intrinsic skyrmion
Hall angle.
We find a multiple step depinning process in which the soliton depins first,
followed next by the depinning of the skyrmion chain along the weak pinning line,
and finally the depinning of the entire skyrmion assembly in the direction
transverse to the drive.
This opens a novel method for precise control of skyrmion motion.

\section{Simulation}

\begin{figure}[h]
    \centering
    \includegraphics[width=\columnwidth]{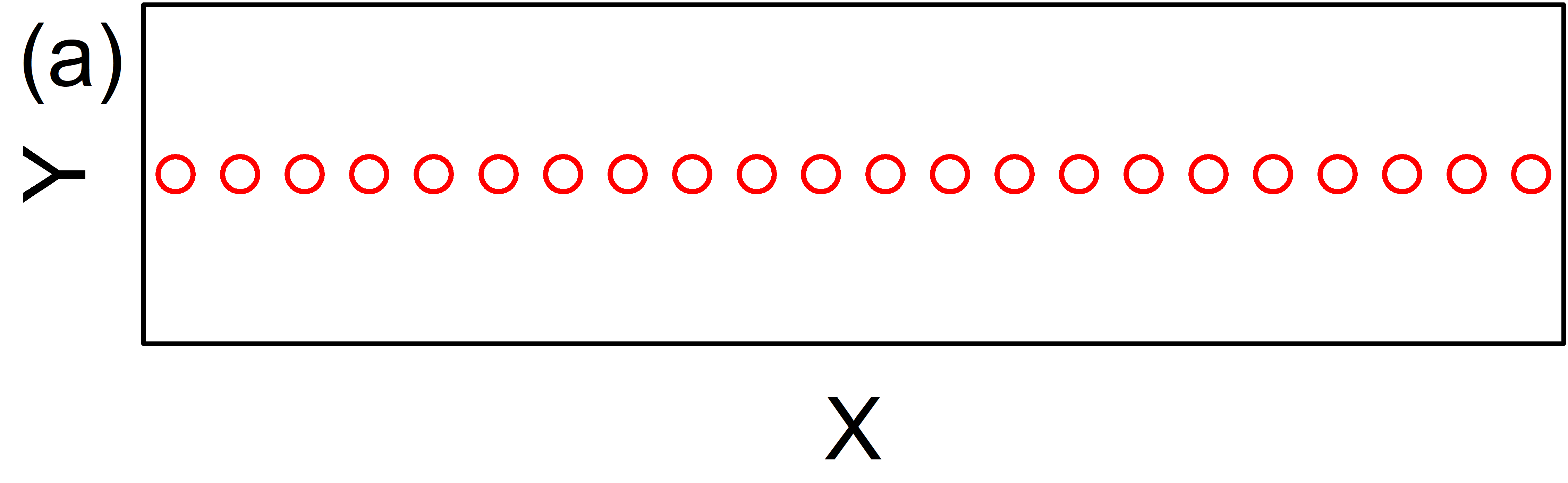}
    \includegraphics[width=\columnwidth]{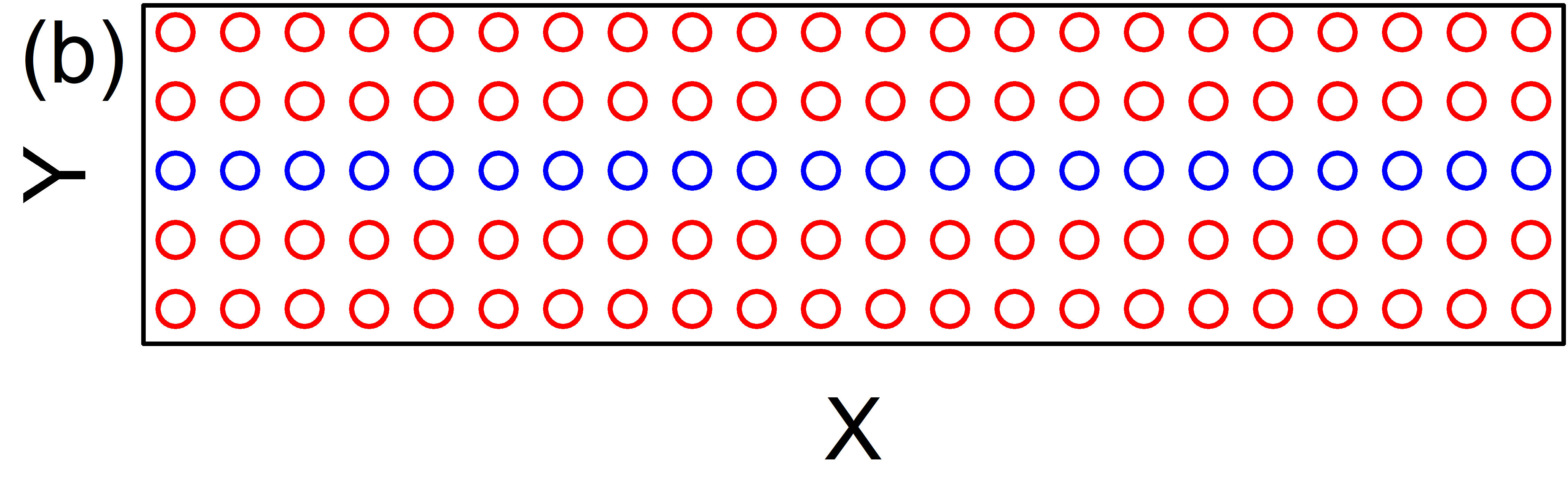}
    \caption{
    Illustration of the samples used in this work. (a) The quasi-one dimensional system where 
the skyrmion motion is confined to a line of Gaussian pinning potentials (circles)
 by  repulsive barriers at $y=0$ and $y=L_y$.
(b) The two-dimensional system with a square array of pinning centers and no 
 repulsive barriers. All pinning sites are modeled using Gaussian pinning potentials.
 Red circles indicate stronger pinning centers
 and the blue circles are the weaker pinning potentials.
    }
    \label{Fig1}
\end{figure}

We simulate the collective behavior of $N_{sk}$ skyrmions interacting with $N_p$ attractive pinning 
centers in a $L_y\times L_x$ two-dimensional
box with periodic boundary 
conditions in both the $x$ and $y$ directions,
as illustrated in Fig.~\ref{Fig1}. The skyrmion density is $n_{sk}=N_{sk}/L_y L_x$ and 
the pinning density is $n_{p}=N_{p}/L_y L_x$. The 
simulations are performed just outside the commensuration ratio 
$N_{sk}/N_p=1$ for either an interstitial skyrmion ($N_{sk} = N_{p}+1$) or a vacancy 
($N_{sk}=N_{p}-1$).
Initially we consider the simplest quasi-one dimensional
case where the skyrmions are confined in a line of Gaussian pinning sites
by repulsive barrier walls
located at $y=0$ and $y=L_y$, as illustrated in Fig.~\ref{Fig1}(a). 
We next work with a sample containing
no repulsive barrier walls where there is
a square lattice of pinning centers bisected by a
line of weaker pinning potentials, as shown in Fig.~\ref{Fig1}(b).
The weak pinning line is aligned with the driving direction in most of this
work, but we also consider the case where the weak pinning is at
$45^\circ$ to the driving direction.

The skyrmion dynamics is governed by
the following particle based equation of
motion \cite{lin_particle_2013}:

\begin{equation} \label{eq:1}
      \alpha_d\mathbf{v}_{i}+\alpha_m\hat{z}\times\mathbf{v_{i}}=\mathbf{F}_{i}^{ss}+\mathbf{F}_{i}^{p}+\mathbf{F}_{i}^{W}+\mathbf{F}^{D} \ .
\end{equation}

The first term on the left hand side represents the damping
that arises from the spin precession and dissipation of electrons
in the skyrmion core, where
$\alpha_d$ is the damping constant. 
The second term on the left hand side is
the Magnus force, where $\alpha_m$ is the Magnus constant.
The Magnus force is oriented perpendicular to
the skyrmion velocity.
The skyrmion-skyrmion repulsive interaction is $\mathbf{F}_i^{ss}= \sum_{i}^{N_{sk}} K_{1} 
(r_{ij}/\xi) {\mathbf{\hat{r}}_{ij}}$, where the screening length $\xi$ is set to $\xi=1.0$
in
dimensionless units, $r_{ij}=|\mathbf{r}_i - \mathbf{r}_j|$ is the distance between skyrmions $i$ and 
$j$, and ${\mathbf{\hat{r}}_{ij}}=(\mathbf{r}_i - \mathbf{r}_j)/r_{ij}$. For better computational 
efficiency, we cut off the
exponentially decaying skyrmion-skyrmion interaction beyond $r_{ij}=6.0$.
We model the interaction between the skyrmions and the pinning centers using the Gaussian form 
$U_p=-C_p e^{(r_{ip}/a_p)^2}$, where $C_p$ is the strength of the pinning potential. Thus, the 
skyrmion-pinning interaction is given by 
$\mathbf{F}_{i}^{p} = -\nabla U_p = F_{p}r_{ip}e^{(r_{ip}/a_p)^2} \mathbf{\hat{r}}_{ip}$, 
where $F_{p}=2C_{p}/a_{p}^2$, $r_ip$ is the distance between skyrmion $i$ and pinning center $p$, 
and $a_p$ is the pinning center radius. In this work we use two types of pinning centers: strong pinning 
centers
with $U_p= 1.0$ and weak pinning centers
with $U_p=0.15$. In both cases the pinning radius is fixed to $a_p=0.3$. We cut off 
this interaction beyond $r_{ip}=2.0$ for computational efficiency. 
The third term on the right hand side, $\mathbf{F}_{i}^{W}$, represents the force exerted by
the repulsive barrier walls.
In
the presence of the barrier walls,
the skyrmion behavior is similar to what would be observed in a
quasi-1D potential well.
The wall potential is
$U_W=U_{W_0}\cos(wy)$, 
where $U_{W_0}=12.0$ and $w=2\pi/L_y$.
The force exerted by the wall is given by $\mathbf{F}_{i}^{W}=-\nabla U_W = -F_{W}\sin(wy)$, where
$F_W = 2\pi U_{W_0}/L_y$.
The
term $\mathbf{F}^{D}=F^{D} \mathbf{\hat{x}}$ represents the applied dc drive,
which is fixed to be along the positive $x$ direction.
We increase $F^{D}$ in small steps of $\delta F^{D}= 0.01$
and spend $2\times 10^{5}$
simulation time steps at each drive increment.
We measure the 
average velocities $\left\langle V_x\right\rangle=\left\langle \mathbf{v} \cdot \widehat{\rm 
{\bf{x}}}\right\rangle$ and $\left\langle V_y\right\rangle = \left\langle \mathbf{v} \cdot
\widehat{\rm {\bf{y}}}\right\rangle$.
We normalize all distances by
the screening length $\xi$ and select
the damping and Magnus constants
such that ${\alpha_m}^2+{\alpha_d}^2=1$.

\section{The Quasi-one dimensional system}

\begin{figure}[h]
   \centering
   \includegraphics[width=1.0\columnwidth]{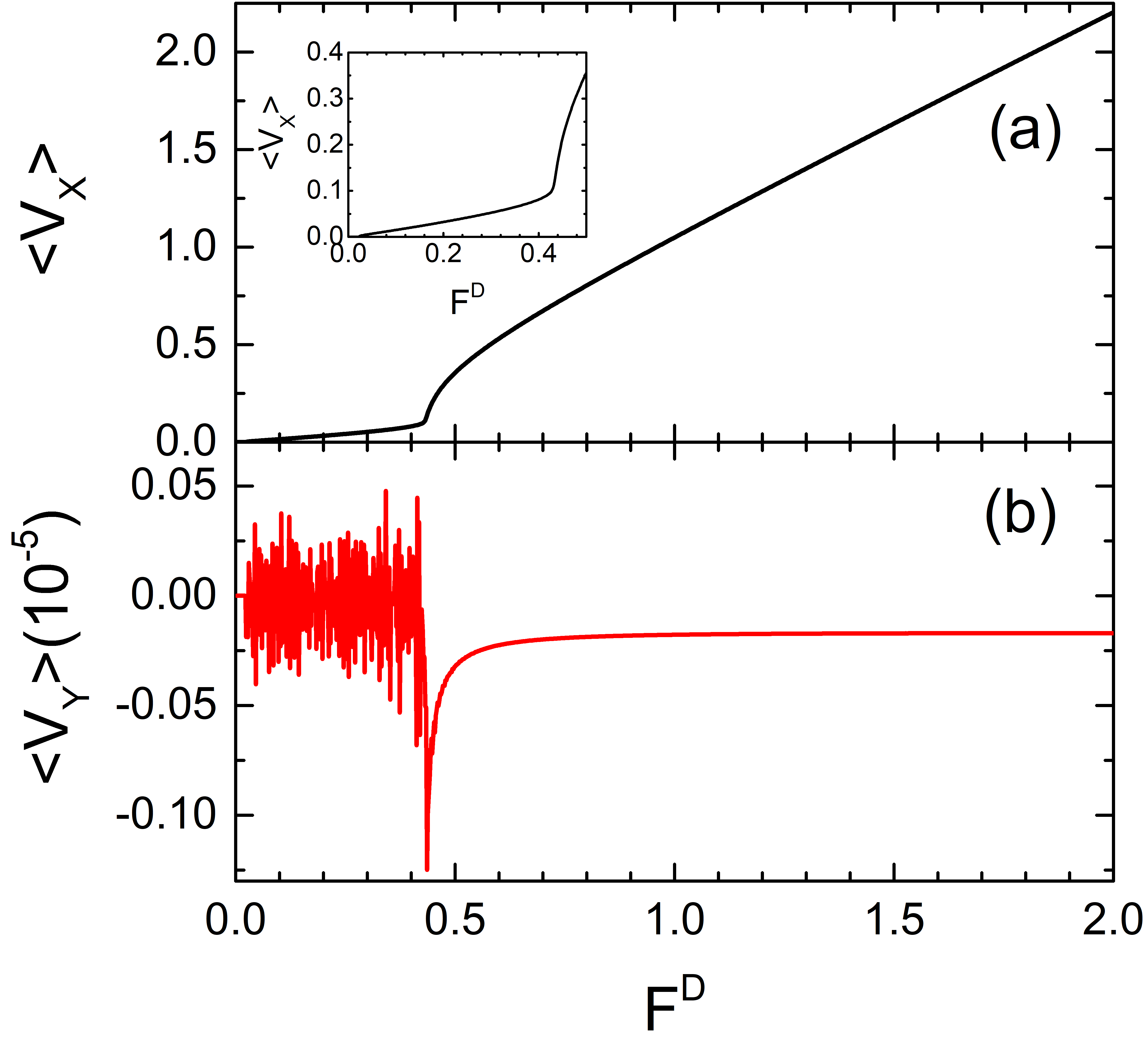}
   \caption{
    (a) $\langle V_x\rangle$ and (b) $\langle V_y\rangle$
     versus the external dc drive $F^D$ for the
     sample illustrated in Fig.~\ref{Fig1}(a) with $N_{sk}/N_p = 1.044$, $\alpha_m/\alpha_d = 0.5$
     and $\rho_p = 0.093\xi^2$.
     The inset of (a) shows a blowup of panel (a)
     over the range $0<F^{D}<0.5$.
 }
   \label{Fig2}
\end{figure}
    
\begin{figure}[h]
  \centering
  \includegraphics[width=1.0\columnwidth]{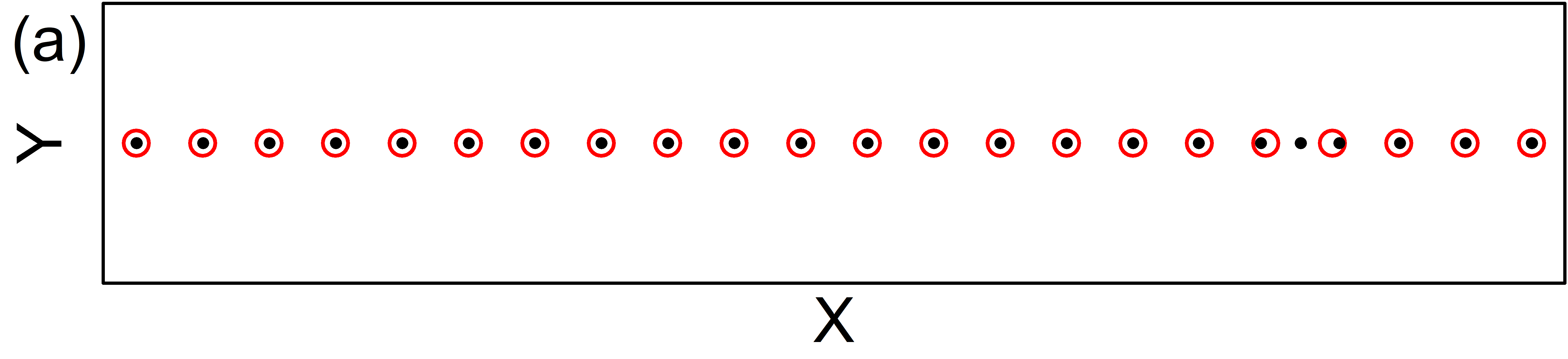}
  \includegraphics[width=1.0\columnwidth]{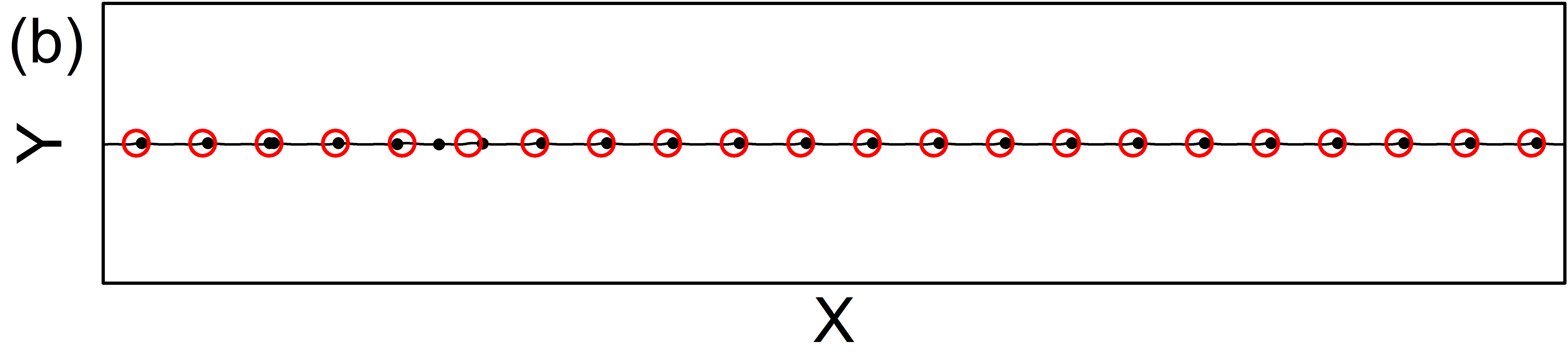}
  \includegraphics[width=1.0\columnwidth]{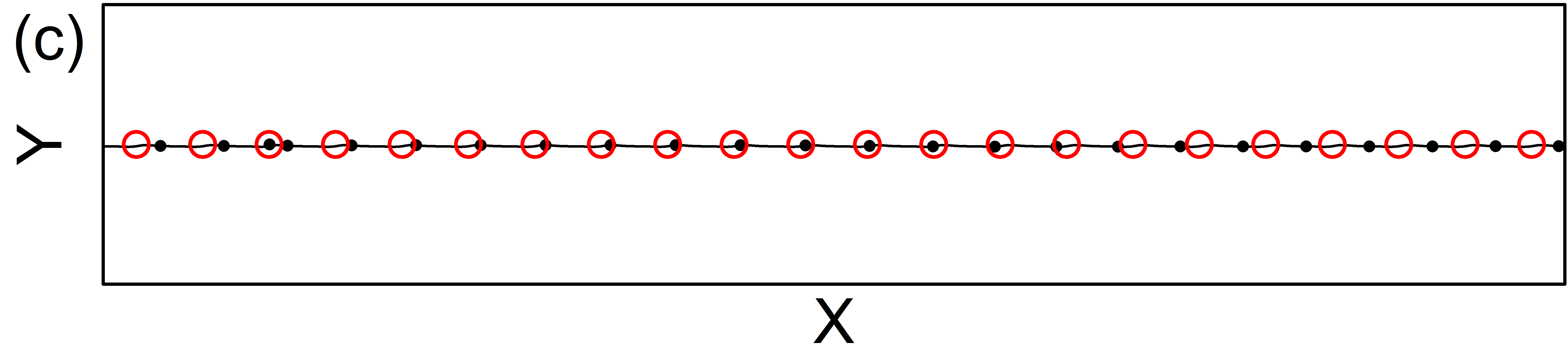}
  \includegraphics[width=1.0\columnwidth]{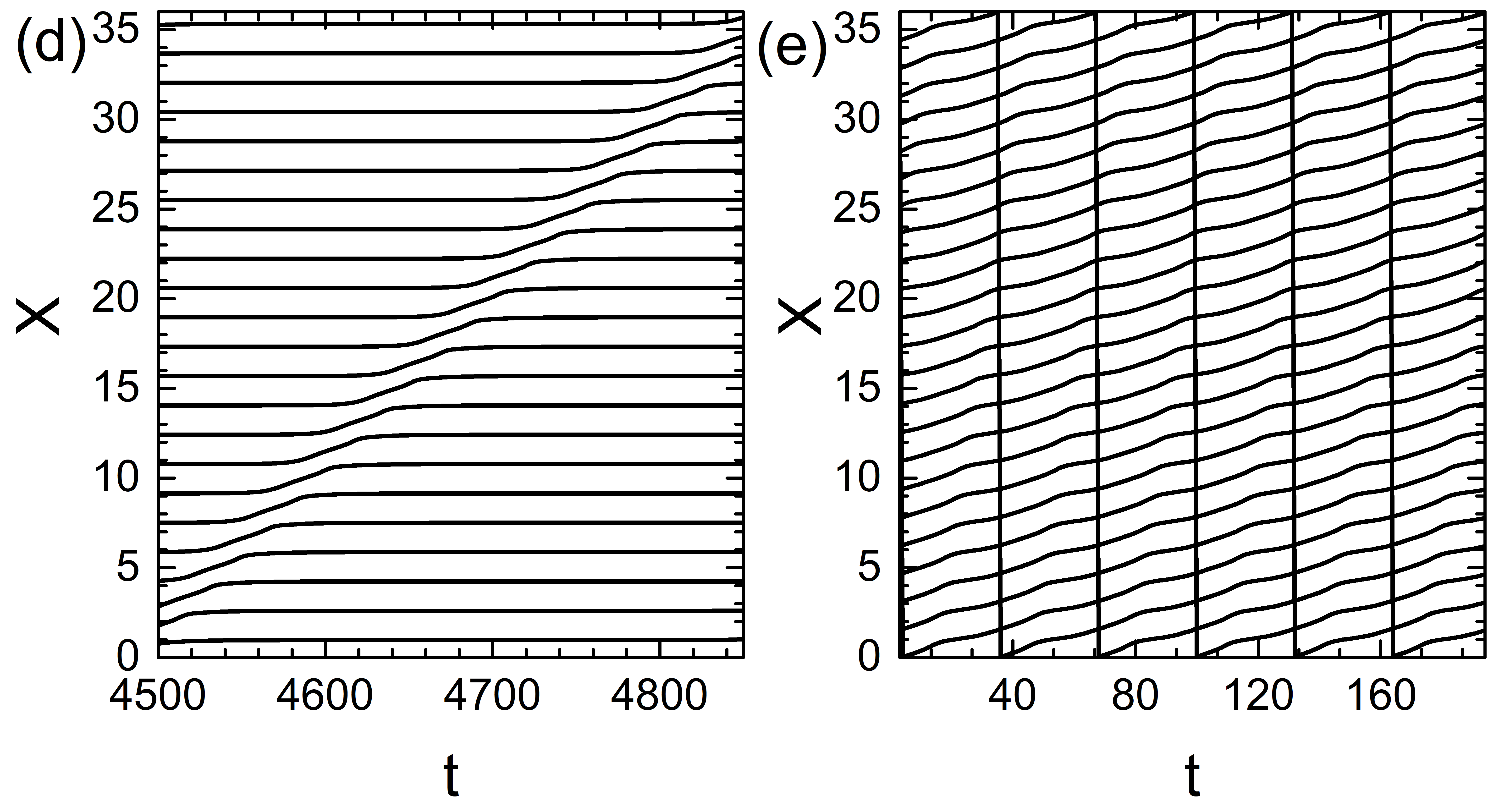}
  
\caption{
(a, b, c) Pinning site positions (red circles) and the skyrmion trajectory (black lines)
for a sample with $N_{sk}=23$, $N_{p}=22$,
$N_{sk}/N_p = 1.044$,
$\alpha_m/\alpha_d = 0.5$, $U_p = 0.15$,
and $\rho_p = 0.093\xi^2$.
(a)
At $F^D=0.01$, the skyrmions are static in the pinned phase.
The incommensuration produces
a deformation in the lattice
in the form of an interstitial skyrmion.
(b) At $F^D = 0.3$,
the interstitial skyrmion moves
as a soliton by hopping from site to site with slow average velocity.
(c)
At $F^D = 1.0$, all of the skyrmions are flowing simultaneously at a higher velocity.
(d, e) Skyrmion positions as a function of time.
(d)
For $F^D=0.3$ as in panel (b),
a soliton pulse propagates through the sample.
(e) For $F^D=1.0$, as in panel (c), all of the skyrmions are flowing in unison.
}
    \label{Fig3}
\end{figure}

We first consider the quasi-one dimensional system illustrated in Fig.~\ref{Fig1}(a).
In this case,
repulsive barrier walls
at $y=0$ and $y=L_y$ surround
an isolated line of
$N_p=22$ weak pinning centers 
filled with $N_{sk}=23$ skyrmions,
giving a value $N_{sk}/N_p = 1.044$ that is just outside a commensurate ratio.
The pinning density is fixed to $\rho_p = 0.093/\xi^2$.
In Fig.~\ref{Fig2} we plot $\langle V_x \rangle$ and $\langle V_y \rangle$ as a
function of the applied dc drive $F^D$ for a system with $\alpha_m/\alpha_d = 0.5$. 
After the depinning
at $F^D = 0.02$, there is a low velocity regime in which
$\langle V_y\rangle$ is noisy and
$\langle V_x \rangle$ increases smoothly and monotonically 
with increasing drive.
The behavior of $\langle V_x\rangle$ is highlighted in the inset of Fig.~\ref{Fig2}(a).
The motion is largely confined to the $x$ direction
by the repulsive 
barrier walls, while the motion in
the $y$ direction is either absent or composed of small amplitude oscillations.
Over the range $0.02<F^D<0.43$,
a soliton pulse is translating
along the skyrmion chain.
Under application of an external drive,
the initial interstitial skyrmion shown in Fig.~\ref{Fig3}(a)
displaces its neighboring skyrmion from the pinning site. The neighboring skyrmion
becomes the new interstitial skyrmion and the previous interstitial skyrmion is
now pinned. The result is a propagation of the location of the interstitial skyrmion
along the chain in the $+x$ direction,
as illustrated in Fig.~\ref{Fig3}(b).
For $F^D > 0.428$,
all of the skyrmions
depin and begin to move collectively,
producing a spike in the velocity-force curve
as shown in Fig.~\ref{Fig2}(a). Due to the 
orderliness of the motion,
the velocity component $\langle V_y \rangle$ drops to zero.

It is difficult to see the differences in motion
between Figs.~\ref{Fig3}(b) and \ref{Fig3}(c) from the overlapping skyrmion
trajectories, so in Fig.~\ref{Fig3}(d,e) we plot the position of each skyrmion as a
function of time.
In Fig.~\ref{Fig3}(d), the system from Fig.~\ref{Fig3}(b) at $F^D=0.3$ contains a
clearly propagating soliton pulse. 
In contrast, Fig.~\ref{Fig3}(e) shows the system from Fig.~\ref{Fig3}(c) at $F^D= 1.0$,
where all of skyrmions are moving coherently as a crystal
and the soliton motion is lost.

\begin{figure}[h]
    \centering
    \includegraphics[width=1.0\columnwidth]{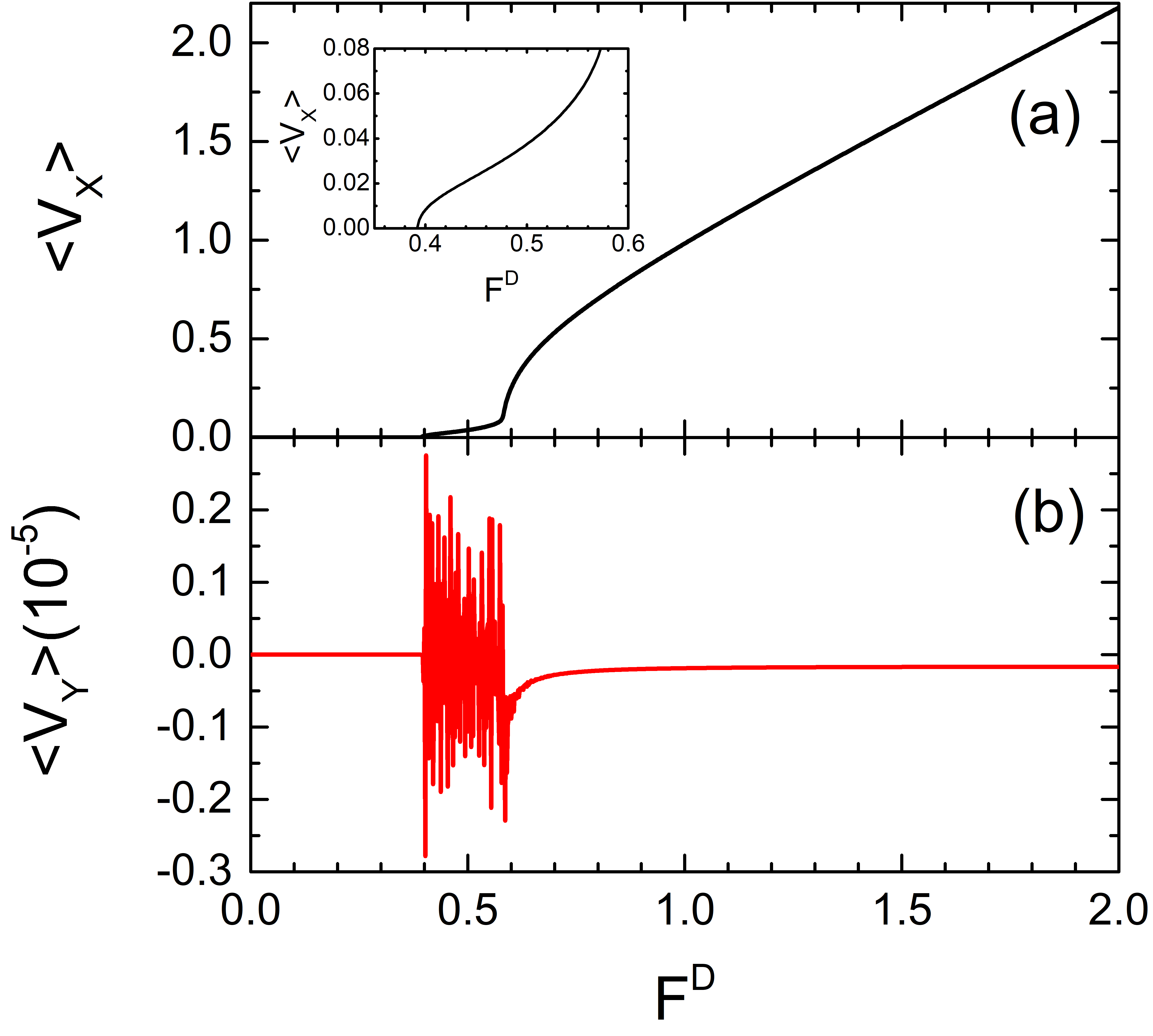}
    \caption{
(a) $\langle V_x\rangle$ and (b) $\langle V_y\rangle$
      versus
      $F^D$ for the
      sample illustrated in Fig.~\ref{Fig1}(a)
      with $N_{sk}/N_p = 0.96$, $\alpha_m/\alpha_d = 0.5$ and 
     $\rho_p = 0.093\xi^2$.
     The inset of (a) shows a blowup of panel (a)
     over the range $0.35<F^{D}<0.55$.
    }
    \label{Fig4}
\end{figure}

\begin{figure}[h]
  \centering
  \includegraphics[width=1.0\columnwidth]{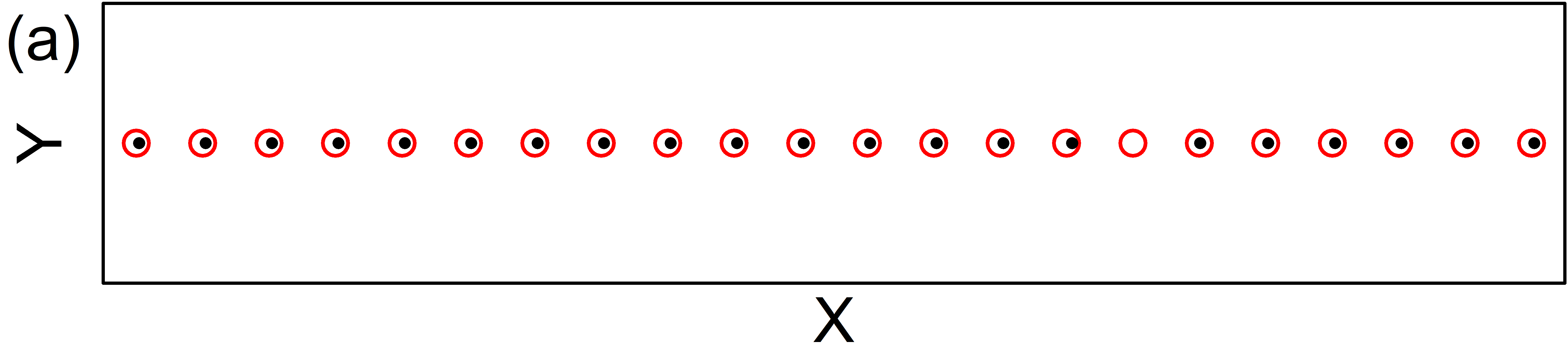}
  \includegraphics[width=1.0\columnwidth]{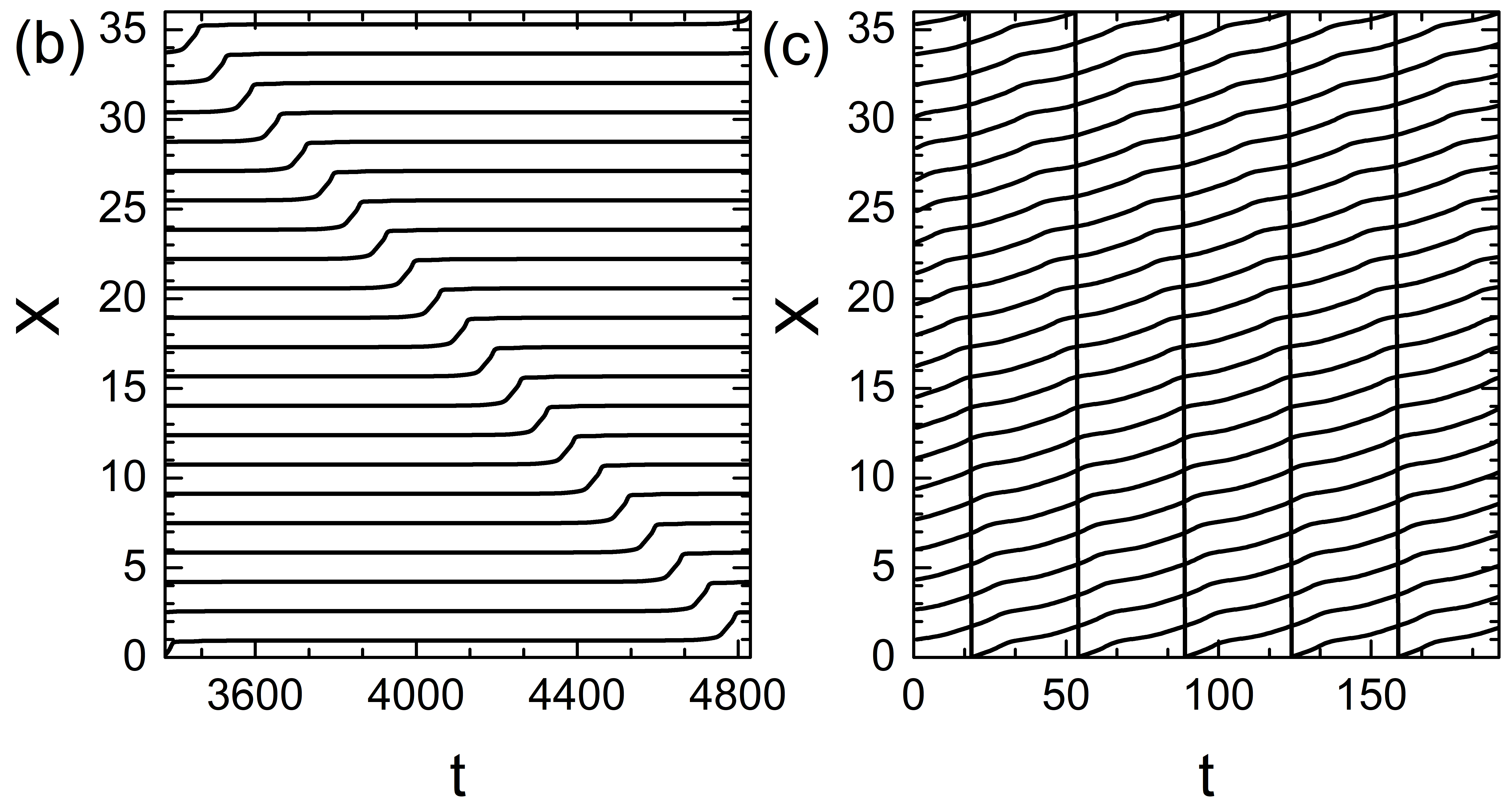}
  
\caption{
(a) Pinning site positions (red circles) and the skyrmion trajectory (black lines)
for a sample with $N_{sk}=21$, $N_p=22$, $N_{sk}/N_p = 0.96$,
$\alpha_m/\alpha_d = 0.5$, $U_p = 0.15$ and $\rho_p = 0.093\xi^2$.
(a) At $F^D=0.3$, the skyrmions are static in the pinned phase.
The incommensuration produces a deformation in the lattice in the form of a
vacant pinning site.
(b, c) Skyrmion positions as a function of time.
(b) At $F^D=0.45$, a soliton pulse propagates in the $-x$ direction through
the sample.
(c) At $F^D=1.0$, all of the skyrmions are flowing in an ordered lattice.
}
    \label{Fig5}
\end{figure}

In Fig.~\ref{Fig4} we plot $\langle V_x \rangle$ and $\langle V_y \rangle$ versus
$F^D$ for a system with $N_{sk}=21$, $N_{p}=22$,
$N_{sk}/N_p = 0.96$, and $\alpha_m/\alpha_d = 0.5$. 
The depinning falls at $F^D = 0.39$, a higher value than that found in
Fig.~\ref{Fig2} due
to the reduced
density of skyrmions in the sample.
Just above depinning, there is a regime of low average velocity similar
to that observed in Fig.~\ref{Fig2}; however, the dynamics is different.
As is shown in Fig.~\ref{Fig5}(a),
there is a vacancy due to the incommensurate ratio between the skyrmions 
and the pinning centers.
This vacancy moves through the sample in the $-x$ direction as the neighboring pinned 
skyrmion depins and fills in the vacancy,
turning its previous pinning site into a new vacancy.
A repetition of this process produces a soliton propagation through the sample
over the range
$0.39<F^D<0.58$, where $\langle V_y\rangle$ is noisy and
$\langle V_x \rangle$ increases smoothly
with increasing $F_D$ as highlighted in the inset of Fig.~\ref{Fig4}(a).
The soliton can be detected experimentally in the same way as
skyrmions by looking for the variation in the skyrmion spacing.
For $F^D>0.58$, all of the skyrmions depin and flow through the sample as a
moving lattice with an average velocity component
$\langle V_x \rangle$ that increases rapidly with increasing $F^D$ and with
$\langle V_y \rangle \approx 0$.
The plot of the skyrmion positions versus time in
Fig.~\ref{Fig5}(d) at $F^D=0.45$ shows the backwards propagation of the
soliton pulse, while a similar plot in Fig.~\ref{Fig5}(e) at $F^D=1.0$ indicates
that all of the skyrmions are moving in unison through the system and the soliton
pulse has been destroyed.

\section{The 2D system}

We next turn to a fully two-dimensional sample containing
no repulsive barrier walls, so that $F_W = 0$.
The sample contains a square array of $N_p=110$ pinning sites, most of
which have a strong $U_p = 1.0$.
As illustrated in Fig.~\ref{Fig1}(b),
there is a central line of weak pinning centers with $U_p = 0.15$,
which serve as a channel to guide the skyrmion motion.
The pinning density in this section is fixed to $\rho_p = 0.373/\xi^2$.

\begin{figure}[h]
    \centering
    \includegraphics[width=0.88\columnwidth]{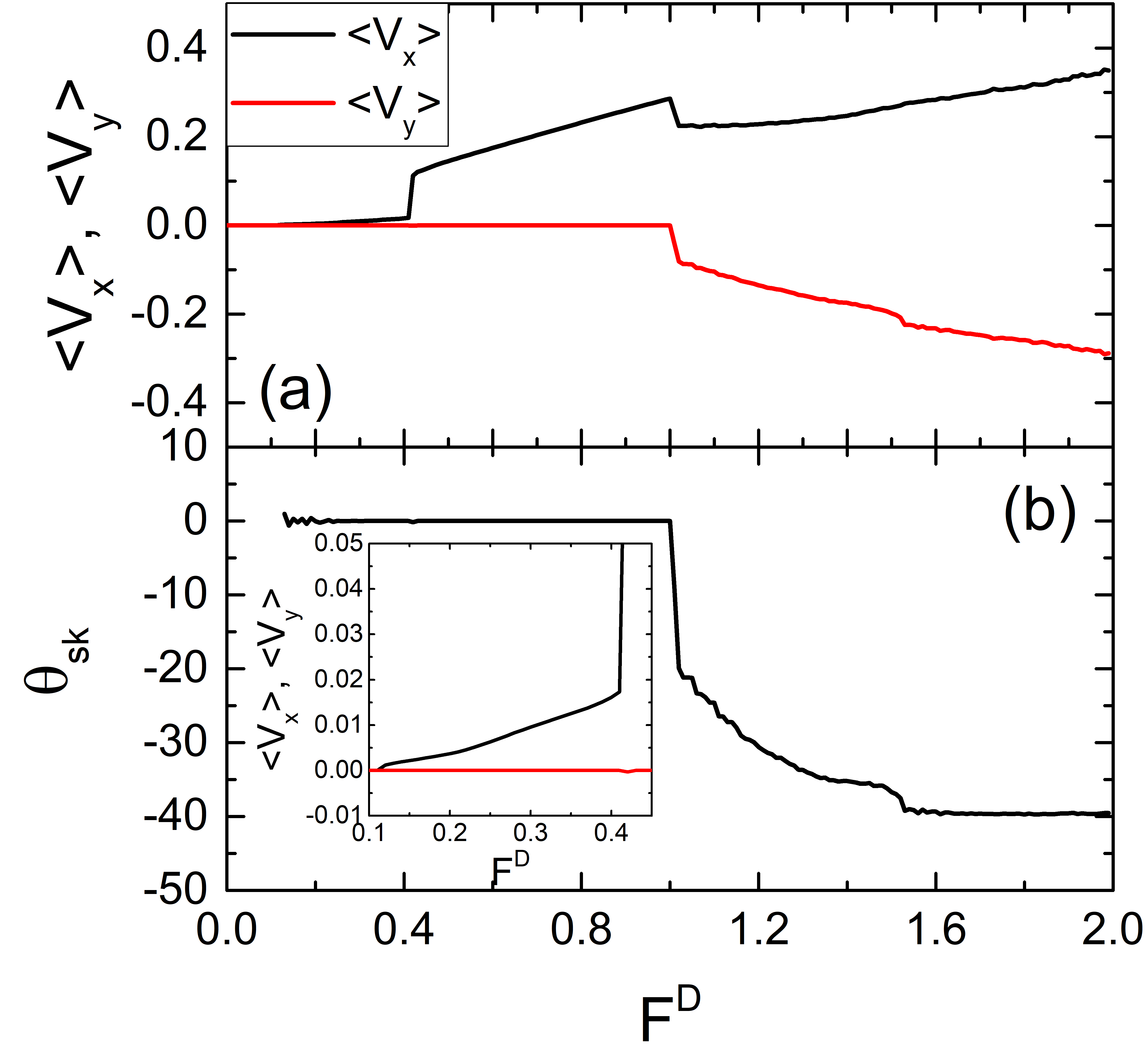}
   \caption{
    (a) $\langle V_x\rangle$ (black) and $\langle V_y\rangle$ (red)
     versus $F^D$ for the 2D
     sample illustrated in Fig.~\ref{Fig1}(b)
     with $N_{sk}/N_p = 1.01$, $\alpha_m/\alpha_d = 1.0$, and 
     $\rho_p = 0.373\xi^2$. Inset: a blowup of panel (a) over the range
     $0.1<F^D<0.45$.
     (b) The corresponding
     skyrmion Hall angle $\theta_{sk}$ versus $F^D$.
    }
    \label{Fig6}
\end{figure}

\begin{figure}[h]
  \centering
    \includegraphics[width=0.9\columnwidth]{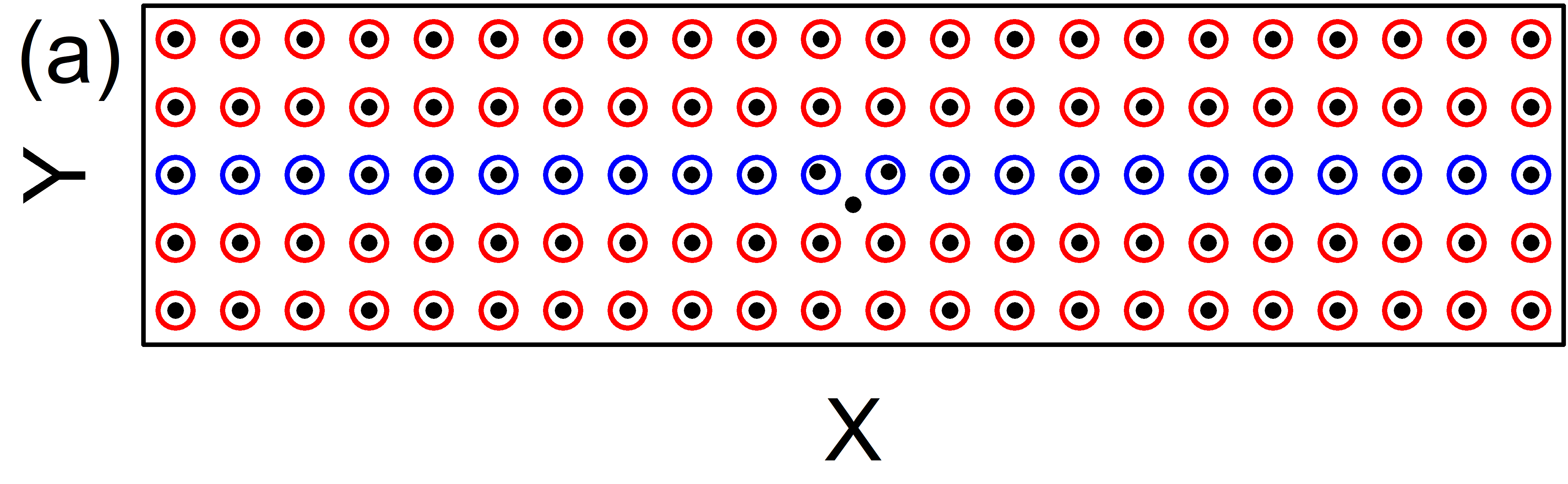}
    \includegraphics[width=0.9\columnwidth]{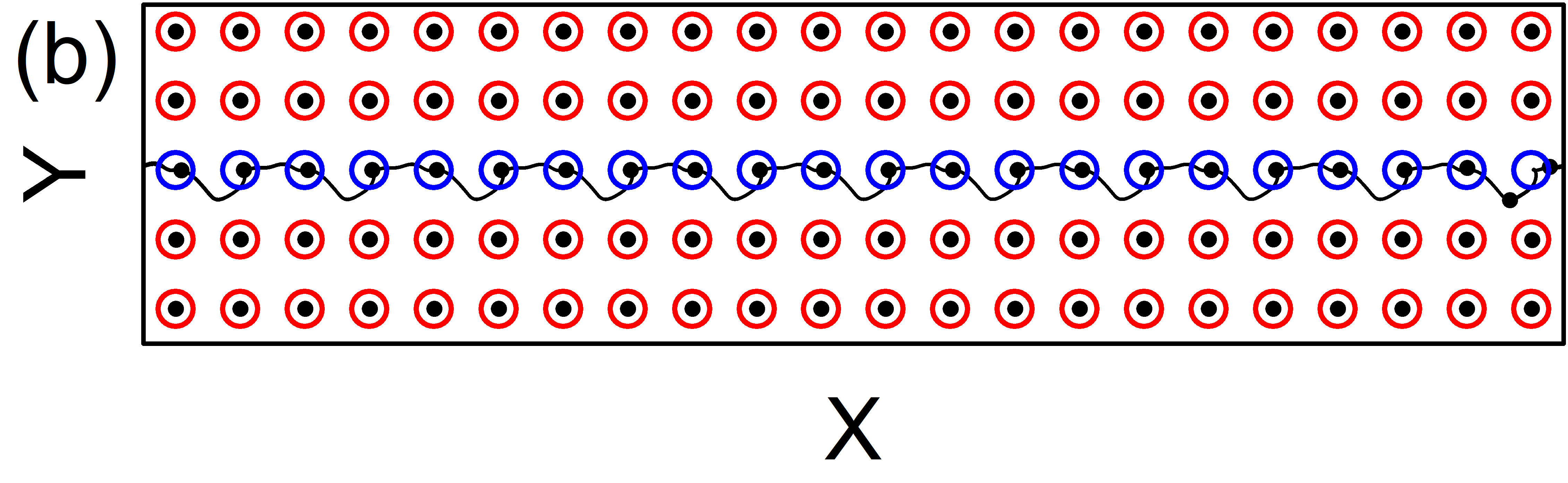}
    \includegraphics[width=0.9\columnwidth]{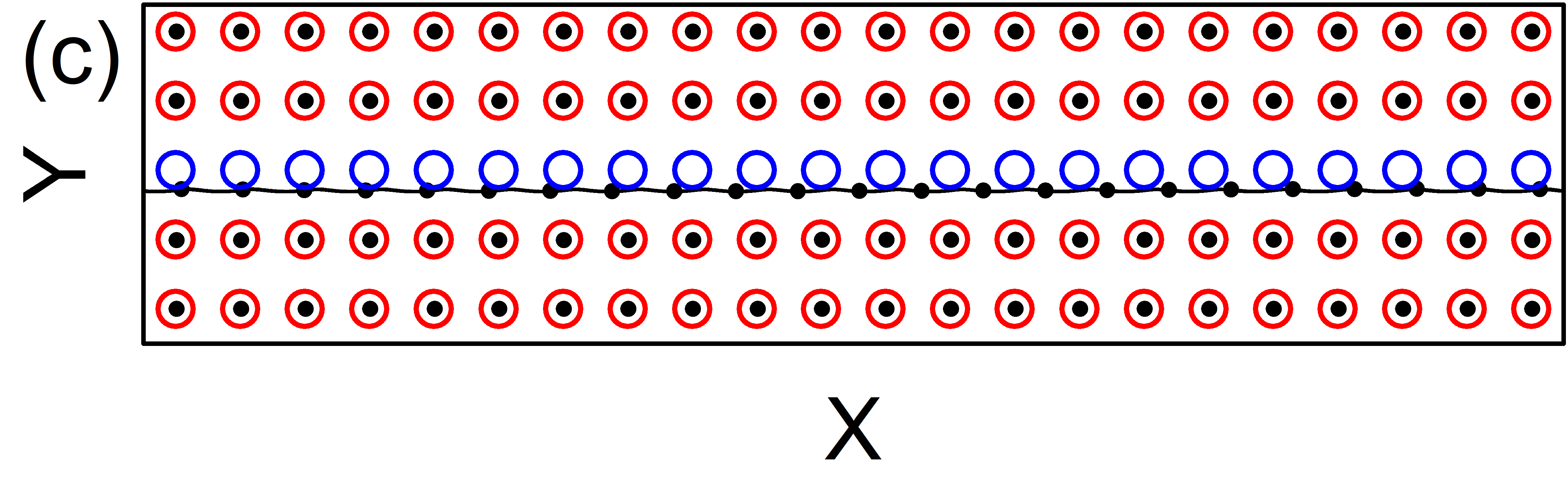}
    \includegraphics[width=0.9\columnwidth]{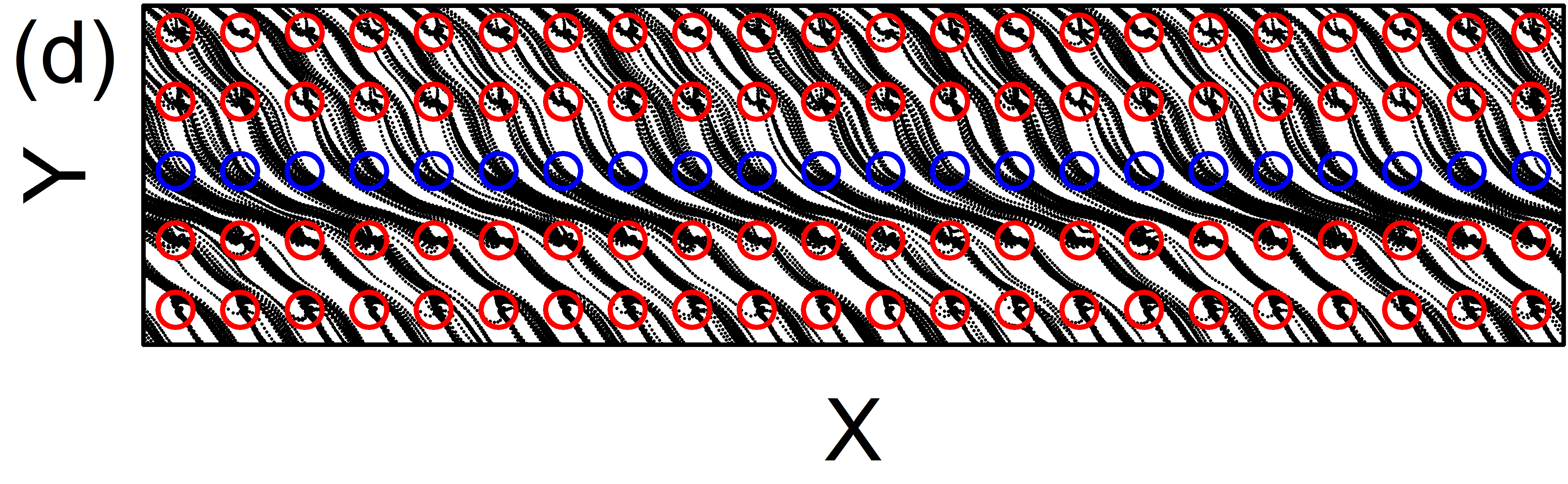}
    \includegraphics[width=0.9\columnwidth]{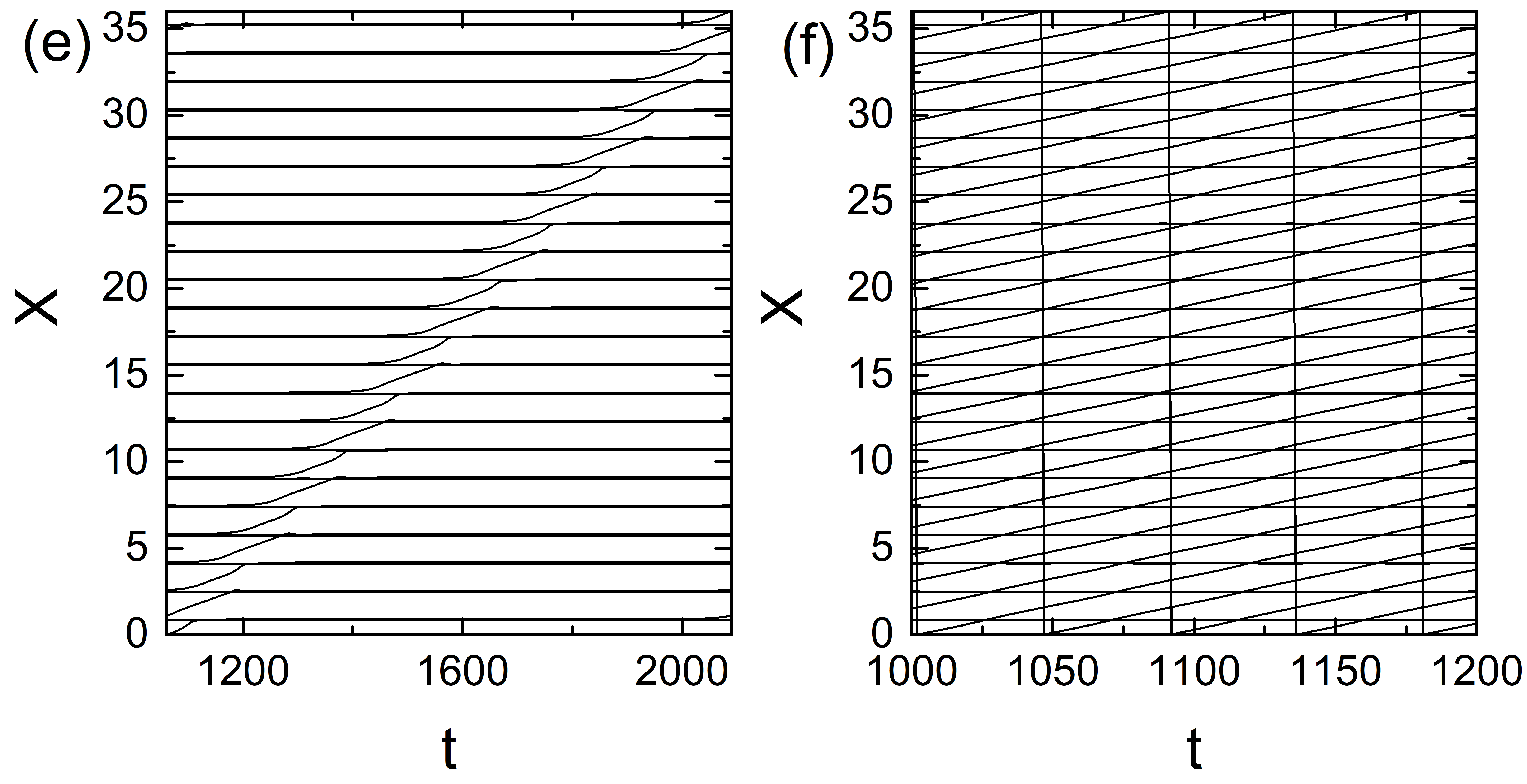}
   \caption{
(a, b, c, d) Pinning site positions (red circles: strong pins; blue circles: weak pins)
and the skyrmion trajectories (black lines)
for a 2D sample with $N_{sk}/N_p = 1.01$,
$\alpha_m/\alpha_d = 1.0$,
weak pins of $U_p = 0.15$, strong pins of $U_p=1.0$, and $\rho_p = 0.373\xi^2$.
(a) $F^D=0$ in the ground state,
where most skyrmions (black circles) are pinned and a single interstitial
skyrmion is present.
(b) The soliton phase at  $F^D = 0.25$.
(c)
At $F^D = 0.5$, all of the skyrmions in the weak pinning centers
depin and flow as a confined chain through the sample.
(d) At $F^D = 1.8$, all of the skyrmions are depinned and flow
along $\theta_{sk}=40^{\circ}$.
(e, f) Skyrmion $x$ positions as a function of time.
(e) The soliton phase at $F^D=0.25$ from panel (b).
(f) The confined chain flow phase at $F^D=0.5$ from panel (c), where the
soliton motion is lost.
}
    \label{Fig7}
\end{figure}

In Fig.~\ref{Fig6}(a) we plot $\langle V_x \rangle$ and $\langle V_y \rangle$ as a
function of the applied dc drive $F^D$
for a system with $N_{sk}=111$, $N_p=110$,
$N_{sk}/N_p = 1.01$, and $\alpha_m/\alpha_d = 1.0$,
while in Fig.~\ref{Fig6}(b) we show
the corresponding skyrmion Hall angle $\theta_{sk}$ versus $F^D$.
The skyrmion dynamics is no longer locked in the $x$ direction, making a diverse
array of dynamic phases possible.
For $F^D \leq 0.11$ the system is in the pinned phase, as 
illustrated in Fig.~\ref{Fig7}(a).
The interstitial skyrmion is localized
between four pinning centers, two of which are strong and two
of which are weak.
The skyrmions trapped in the weaker pinning potentials experience a greater
displacement due to the neighboring interstitial skyrmion.
For $0.11<F^D<0.41$, we find a soliton phase
very similar to that shown in Fig.~\ref{Fig2} and Fig.~\ref{Fig3}(b) for the 
quasi-one dimensional system.
The interstitial skyrmion displaces a skyrmion from a weak pinning site, taking its
place as a pinned skyrmion and turning the formerly pinned skyrmion into the
new interstitial skyrmion.
This process propagates along the chain, resulting in a soliton pulse moving in the
$+x$ direction.
The skyrmion trajectories for this regime are illustrated
in Fig.~\ref{Fig7}(b), which shows that
oscillations in the $y$ direction
occur due to the combination of 
the skyrmion Hall angle effect
and the swapping of interstitial and pinned skyrmions.
For $0.41<F^D<1.0$, all of the skyrmions trapped in the weaker pinning 
potentials depin,
resulting in an almost 1D motion with very small oscillations in $y$, as illustrated in
Fig.~\ref{Fig7}(c).
For $F^D>1.0$, the skyrmions in the stronger pinning potentials also depin, resulting
in a 2D motion.
This motion occurs in
two distinct phases that are visible in Fig.~\ref{Fig6}.
In the chaotic phase, found for $1.0<F^D<1.53$,
the skyrmion Hall angle increases irregularly in magnitude, while
for $F^D>1.53$, the skyrmion Hall angle
stabilizes at $\theta_{sk} \approx -40^{\circ}$.
If the applied drive were increased further,
we expect that $\theta_{sk}$ would approach
the intrinsic Hall angle, which in this case is
$\theta_{sk}^{\rm int}=\arctan{(\alpha_m/\alpha_d)}=-45^{\circ}$.
In Fig.~\ref{Fig7}(d) we plot the skyrmion trajectories for $F^D=1.8$, where 
the skyrmion Hall angle is stabilized, showing an orderly 2D motion.
As in the quasi-1D system, the soliton phase is most easily identified by
plotting the skyrmion positions 
as a function of time.
In Fig.~\ref{Fig7}(e) we plot the $x$ position of the skyrmions
as a function of time
for $F^D=0.25$, where a soliton pulse propagates in the $+x$ direction.
In contrast, for $F^D=0.5$,
Fig.~\ref{Fig7}(f) indicates that the pulsed motion
has been destroyed and the skyrmions move as a confined chain.

\begin{figure}[h]
    \centering
    \includegraphics[width=1.0\columnwidth]{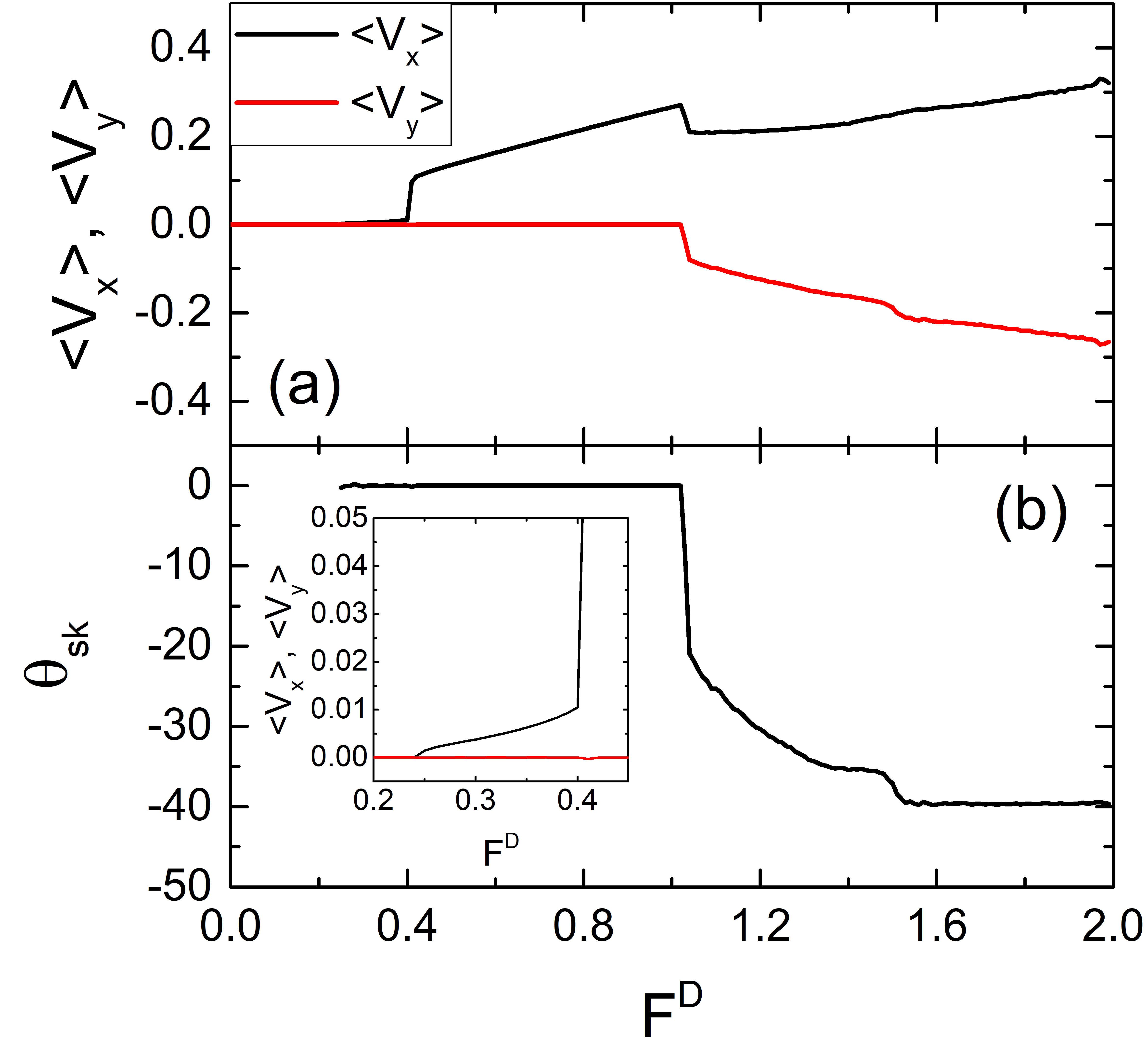}
    \caption{
    (a) $\langle V_x\rangle$ (black) and $\langle V_y\rangle$ (red)
     versus $F^D$ for the 2D
     sample from Fig.~\ref{Fig1}(b) with $N_{sk}/N_p = 0.99$, $\alpha_m/\alpha_d = 1.0$,
     and $\rho_p = 0.373\xi^2$.
     (b) The corresponding skyrmion Hall angle $\theta_{sk}$ versus $F^D$.
     Inset:  a blowup of panel (a)
     over the range $0.2<F^{D}<0.45$.
    }
    \label{Fig8}
\end{figure}

\begin{figure}[h]
      \centering
    \includegraphics[width=0.9\columnwidth]{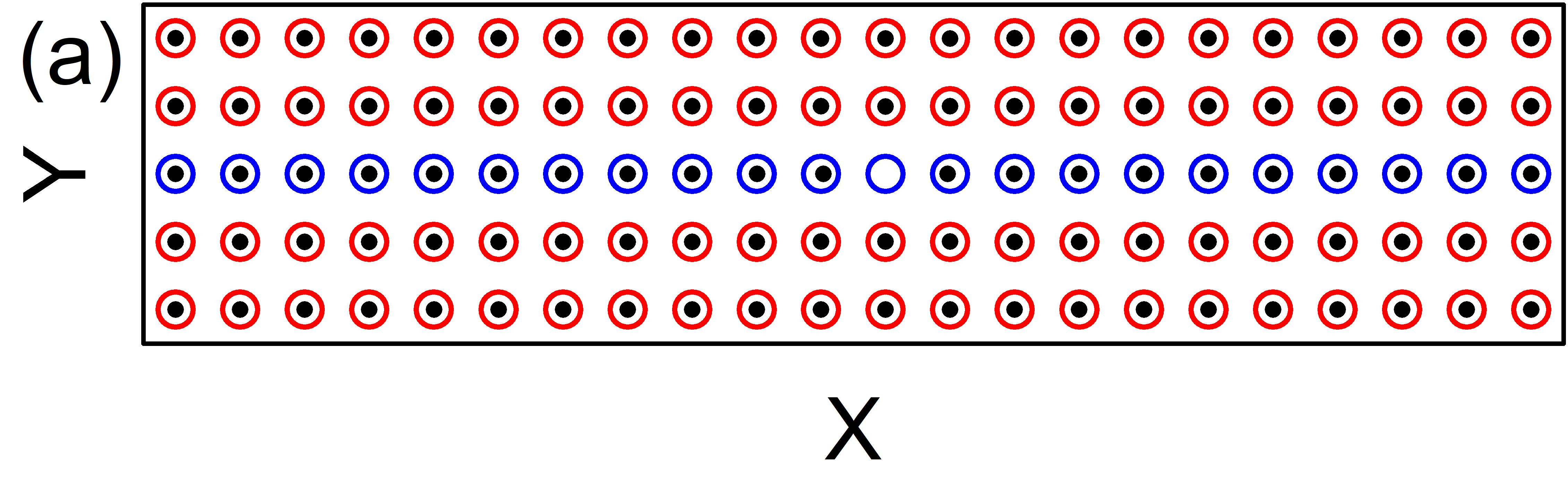}
    \includegraphics[width=0.9\columnwidth]{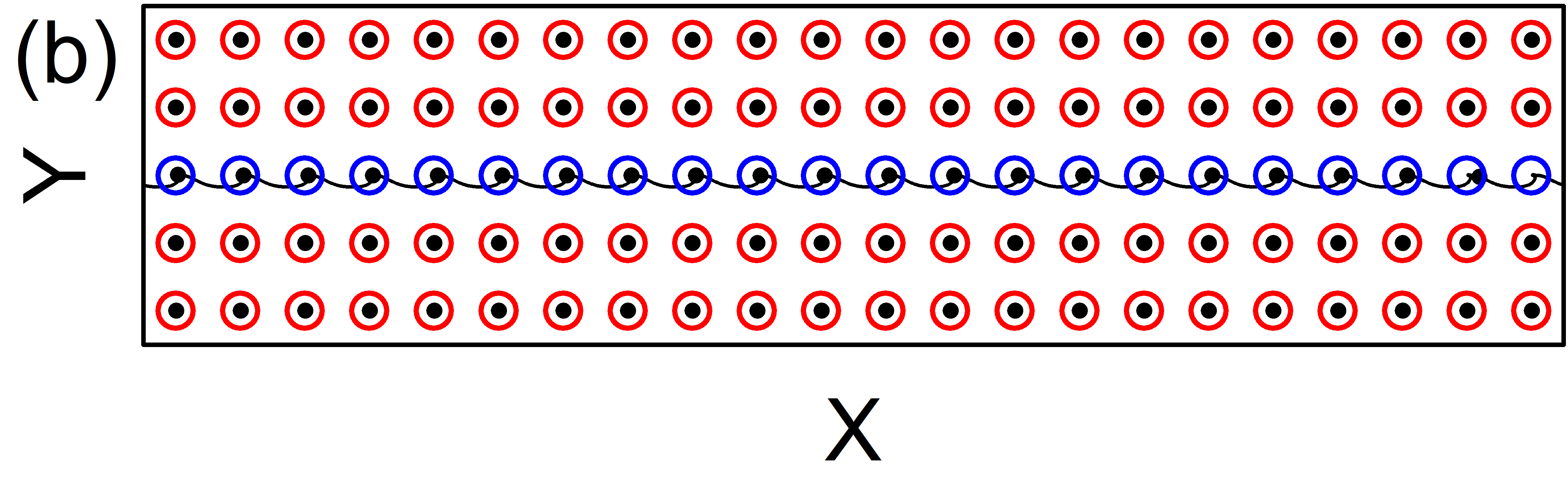}
    \includegraphics[width=0.9\columnwidth]{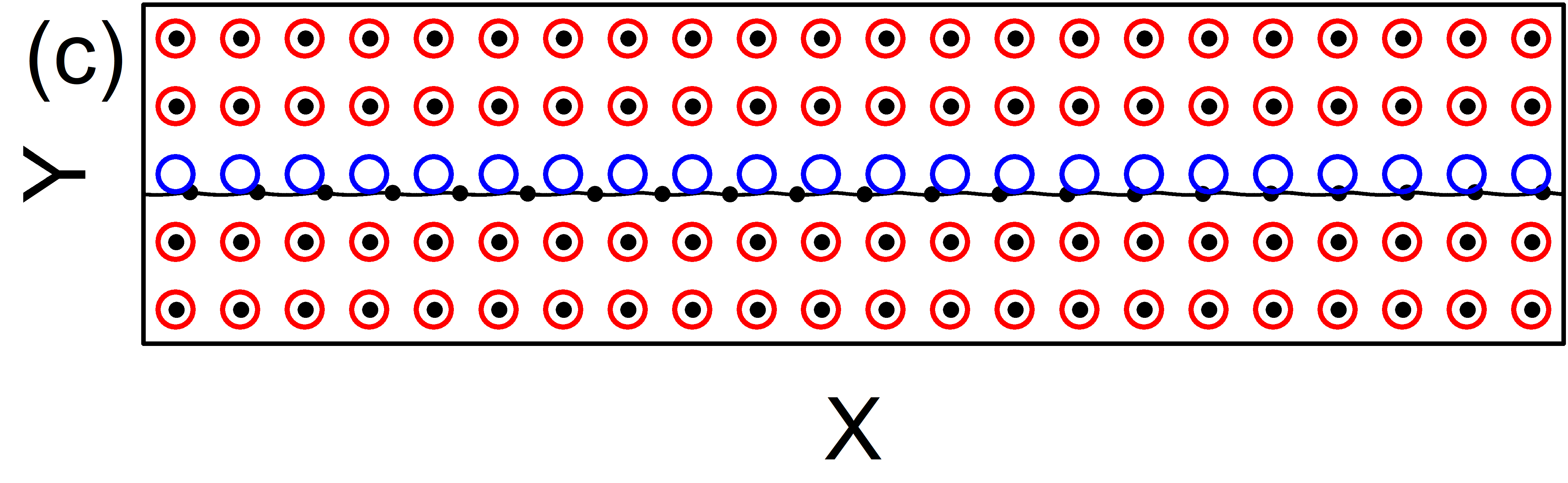}
    \includegraphics[width=0.9\columnwidth]{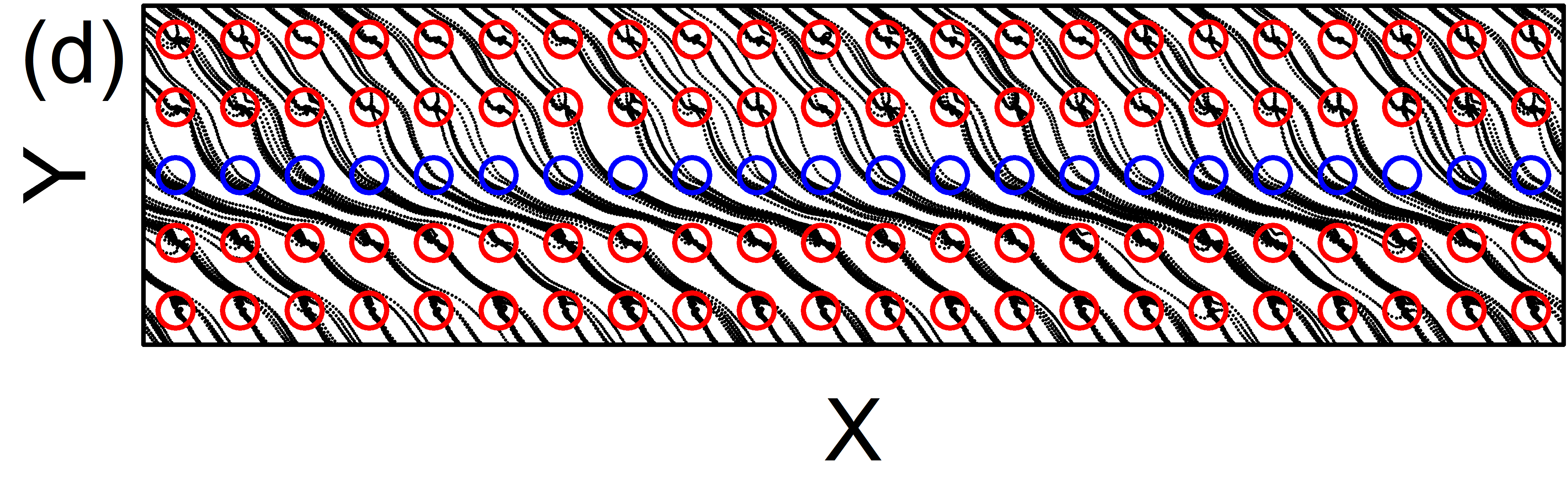}
    \includegraphics[width=0.9\columnwidth]{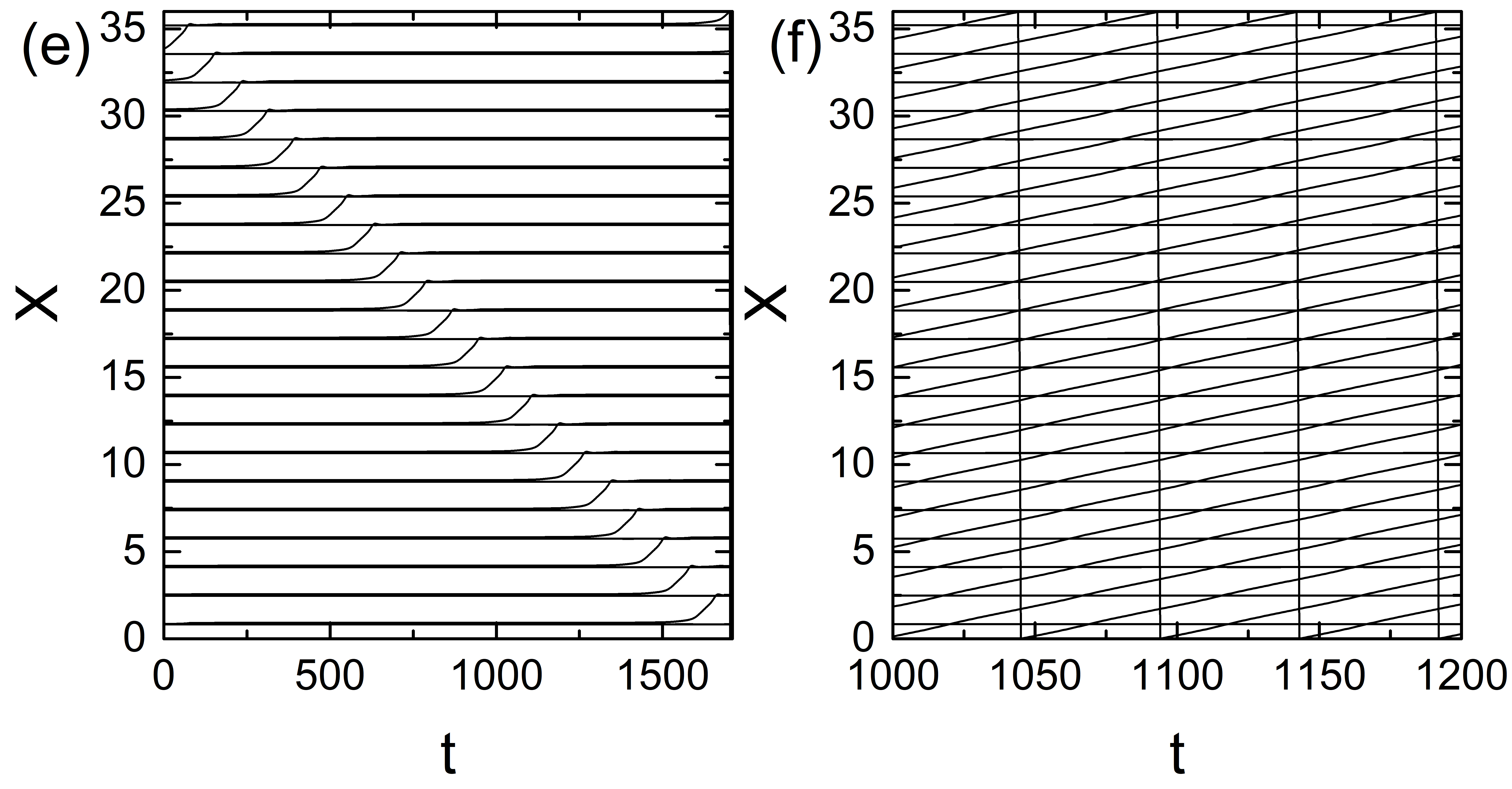}
   \caption{
(a, b, c, d) Pinning site positions (red circles: strong pins; blue circles: weak pins)
and the skyrmion trajectories (black lines)
for a 2D sample with $N_{sk}/N_p = 0.99$,
$\alpha_m/\alpha_d = 1.0$, weak pins of $U_p = 0.15$, strong pins of $U_p=1.0$,
and $\rho_p = 0.373 \xi^2$.
(a)
The ground state at $F^D=0$, where all skyrmions are pinned and
a single vacancy is present.
(b)
The soliton phase at $F^D = 0.3$.
(c)
At $F^D = 0.5$, all
of the skyrmions in the weak pinning centers depin and flow
as a chain through the sample.
(d) 
At $F^D = 1.6$, all of the skyrmions are depinned and flow
with $\theta_{sk}=-40^{\circ}.$
(e, f) Skyrmion $x$ positions as a function of time.
(e) The soliton phase at $F^D=0.3$ from panel (b).
(e) The flowing chain phase at $F^D=0.5$ from panel (c),
where the soliton motion is lost.
}
\label{Fig9}
\end{figure}

In Fig.~\ref{Fig8}(a) we plot $\langle V_x \rangle$ and $\langle V_y \rangle$ as a
function of the applied dc drive $F^D$ for a system with $N_{sk}=109$, $N_p=110$,
$N_{sk}/N_p = 0.99$, and $\alpha_m/\alpha_d = 1.0$, while in Fig. \ref{Fig8} (b) we show
the corresponding skyrmion Hall angle $\theta_{sk}$ versus $F^D$.
For $F^D \leq 0.24$ the system is in the pinned state, as 
shown in Fig.~\ref{Fig9}(a). Due to the incommensurate ratio between the skyrmions 
and the pinning centers, 
there is a vacant pinning center which distorts the lattice. 
For $0.24<F^D<0.4$ the system enters a
soliton phase similar to that shown in Fig.~\ref{Fig4} and Fig.~\ref{Fig5}(b) for the 
quasi-one dimensional case.
The vacancy is pushed in the $-x$ direction due to the hopping motion of
individual skyrmions in
the $+x$ direction.
The trajectories in this regime are illustrated
in Fig.~\ref{Fig9}(b), where small oscillations in the $y$ direction are visible.
For $0.4<F^D<1.02$, all of the skyrmions trapped in the weaker pinning 
potentials depin, resulting in an almost 1D motion with very small oscillations in $y$,
as shown in Fig.~\ref{Fig9}(c). When $F^D>1.02$,
the skyrmions in the stronger pinning centers also depin.
Similarly to what was observed in Fig.~\ref{Fig6},
Fig.~\ref{Fig8} indicates that there are
two dynamic phases for $F^D>1.02$:
a chaotic phase in the
range $1.02<F^D<1.53$, and a more ordered phase
for $F^D>1.53$.
For the latter phase, the skyrmion Hall angle
again stabilizes near $\theta_{sk} \approx -40^{\circ}$.
The similarities between the dynamics of both the interstitial and vacancy
systems at high drives is expected since the difference in the skyrmion density
is very low and becomes unimportant in the drive-dominated regime.
Instead, distinct behaviors
arise in the soliton regime.
To illustrate this,
in Fig.~\ref{Fig9}(e) we plot the skyrmion $x$ positions as a function of time
at $F^D=0.3$, where a moving soliton pulse is clearly visible.
In contrast, for $F^D=0.5$,
Fig.~\ref{Fig9}(f) shows that the pulsed motion
is lost.

The soliton phases for the interstitial and vacancy phases have similar dynamics,
but exhibit the crucial difference that
the interstitial soliton moves in the $+x$ direction while the vacancy
soliton moves in the $-x$ direction.
This interesting behavior, which is stable
over a range of external dc drives, can be harnessed in devices
to allow very low 
external currents
to propagate the soliton through the sample in a fast and controlled manner.

\section{Soliton stabilization as a function of $\alpha_m/\alpha_d$}

\begin{figure}[h]
\centering
\includegraphics[width=1.0\columnwidth]{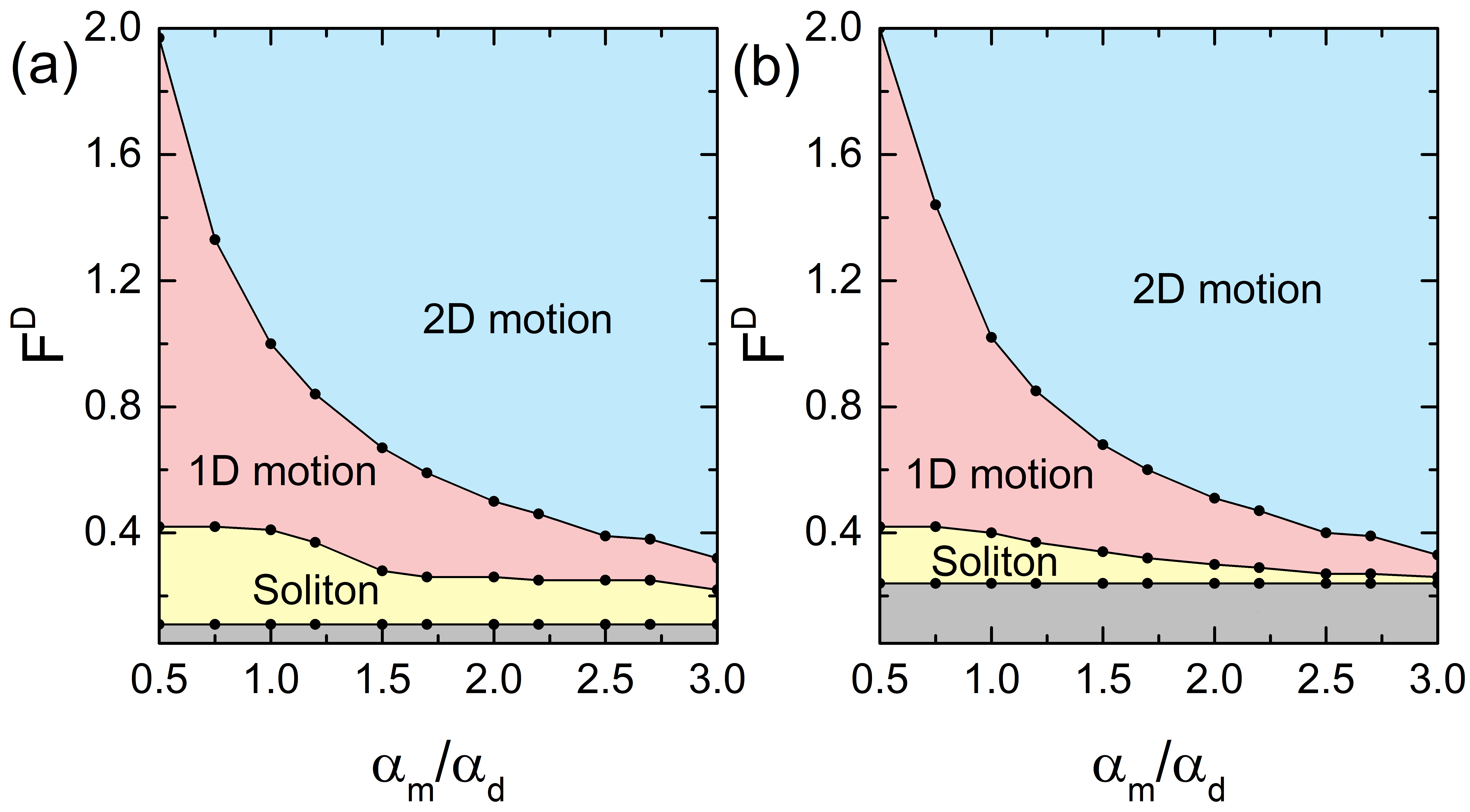}
\caption{
Dynamic phase diagrams
as a function of external dc drive  $F^D$ versus $\alpha_m/\alpha_d$
for the system in Fig.~\ref{Fig1}(b) at $\rho_p = 0.373\xi^2$ and
(a) $N_{sk}/N_p = 1.01$ and (b) $N_{sk}/N_p = 0.99$.
Pinned phase: gray; soliton phase: yellow; 1D motion: red; 2D motion: blue.
}
\label{Fig10}
\end{figure}

We next consider the evolution of the soliton phase as 
$\alpha_m/\alpha_d$ is varied.
When $\alpha_m/\alpha_d$  increases,
the intrinsic skyrmion Hall angle also increases, so it is important
to verify whether the soliton phase remains stable
under these circumstances.
We prepare two samples with
fixed values of $N_{sk}/N_p = 1.01$ and $N_{sk}/N_p = 0.99$ and
perform simulations for a range of 
values of $\alpha_m/\alpha_d$.
By combining the resulting data we generate dynamic phase diagrams as
a function of $F_D$ versus $\alpha_m/\alpha_d$, shown in
Fig.~\ref{Fig10},
where we identify the locations of
the pinned phase,
the soliton phase,
1D chain motion, and 2D motion.
The pinned phase is a static state in which
all pinned skyrmions remain
trapped in the pinning centers and $\langle V_x\rangle=\langle V_y\rangle=0.$
In the soliton phase,
the localized lattice deformation
propagates through the sample.
This soliton travels in the $+x$ direction when
interstitial skyrmions are present and  in the $-x$ direction
when vacancies are present.
1D motion
occurs when all of the skyrmions trapped in the weak pinning potentials depin
and flow as a coherent chain in the $+x$
direction.
In 2D motion,
all of the skyrmions in all of the pinning sites depin and
flow through the sample along both the $x$ and $y$ directions.

At $N_{sk}/N_p = 1.01$, Fig.~\ref{Fig10}(a) indicates that
the depinning threshold is very low, producing a wider range of soliton motion
compared to the system in Fig.~\ref{Fig10}(b)
with $N_{sk}/N_p = 0.99$.
Interstitial skyrmions are more mobile than vacancies since
an interstitial
skyrmion is trapped only by
the caging potentials of the neighboring skyrmions and not directly by a pinning site.
This lowers the depinning threshold for the interstitial system.
Both systems show
a transition from the soliton phase to 1D motion
at roughly the same value of $F_D$ since this transition is controlled by the
strength of the weak pinning sites.
Similarly, the transition line between 1D motion and 2D motion,
which is controlled by the strength of the strong pinning sites,
falls at similar values of $F_D$ in both systems.

\section{Effect of pinning density}

\begin{figure}[h]
    \centering
    \includegraphics[width=1.0\columnwidth]{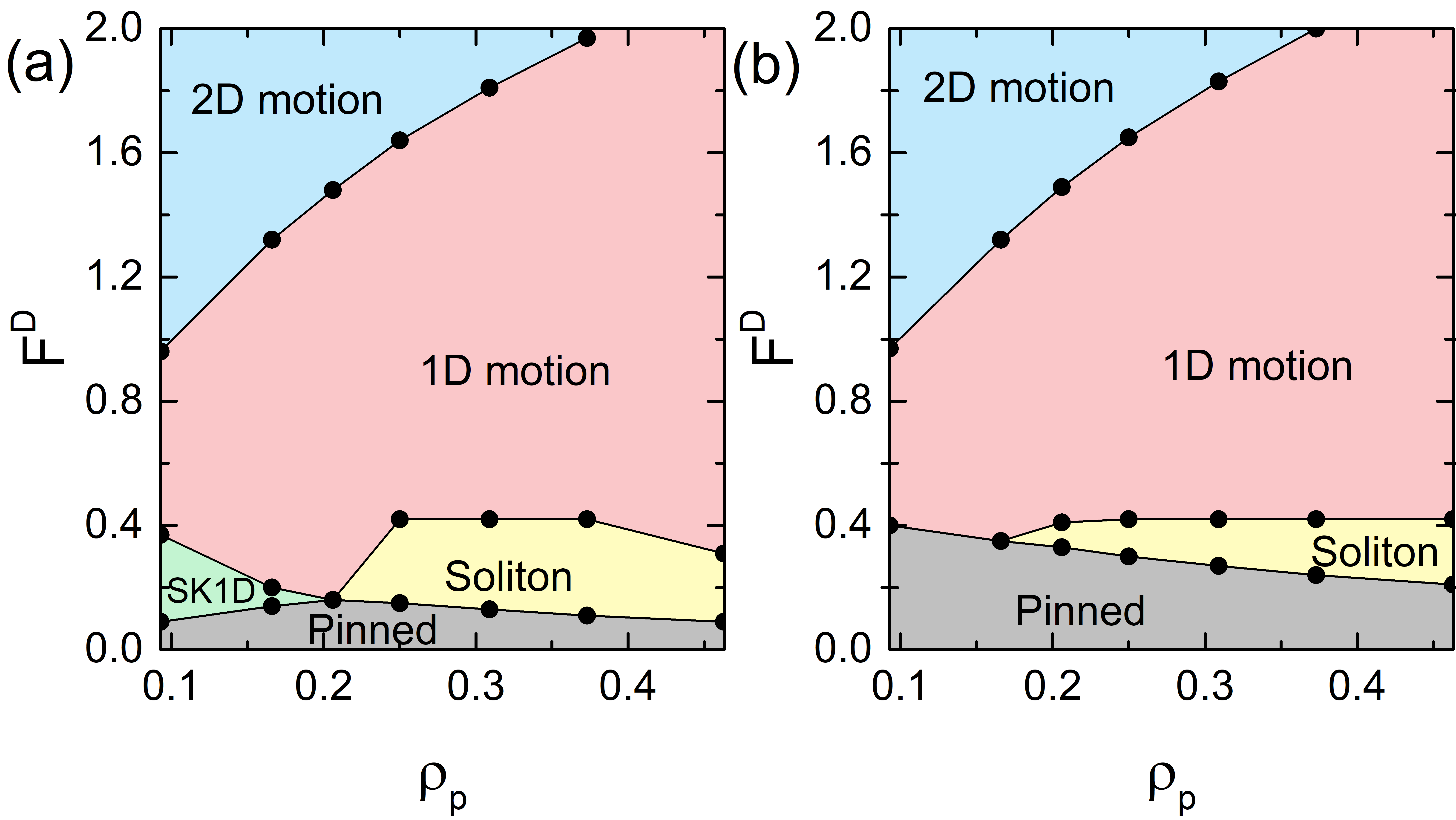}
    \includegraphics[width=0.7\columnwidth]{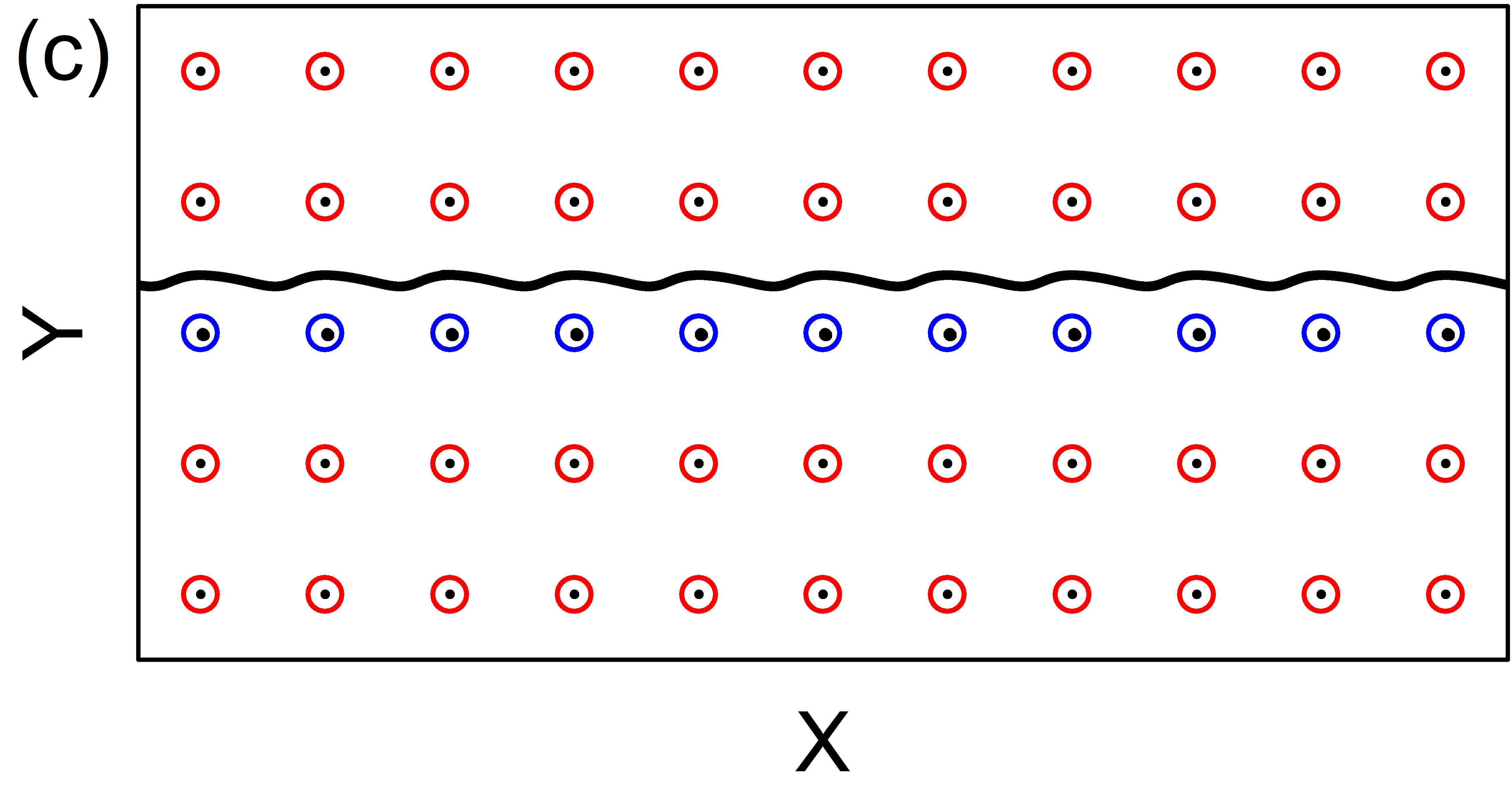}
\caption{
(a, b) Dynamic phase diagrams as a function of $F^D$ versus pinning density
$\rho_p$ for the samples from Fig.~\ref{Fig1}(b) with $\alpha_m/\alpha_d=0.5$,
weak pinning of $U_p=0.15$, and strong pinning of $U_p=1.0$ for  
(a) $N_{sk}/N_p = 1.01$ and (b) $N_{sk}/N_p = 0.99$.
Pinned phase: gray; soliton phase: yellow; 1D motion: red; 2D motion: blue;
single skyrmion 1D motion (SK1D): green.
(c) Pinning site positions (red circles: strong pins; blue circles: weak pins)
and the skyrmion trajectories (black lines)
for the $N_{sk}/N_p = 1.01$ sample at $F^D = 0.2$
and $\rho_p = 0.093\xi^2$.
}
    \label{Fig11}
\end{figure}

We next vary the pinning density $\rho_p$ for
$N_{sk}/N_p = \rho_{sk}/\rho_p = 1.01$
and $0.99$
while fixing $\alpha_m/\alpha_d=0.5$.
When the pinning density is low, we expect that
the soliton motion will
vanish when the large spacing between adjacent pinning sites
destroys
the collective behavior.
In Fig.~\ref{Fig11}(a,b) we show dynamic phase diagrams as a function of
$F^D$ versus $\rho_p$ for both systems.
Here we observe a pinned phase, a soliton phase, 1D motion, 2D motion,
and an additional state that we term single skyrmion 1D motion (SK1D).
In the SK1D state,
the interstitial skyrmion
produced by the incommensuration in the $N_{sk}/N_p=1.01$
sample flows
between the pinning centers without displacing any of the pinned skyrmions.
This behavior occurs
only for low pinning densities, $\rho_p<0.206$, when the gaps between adjacent
pinning sites are sufficiently large,
as shown in Fig.~\ref{Fig11}(c).
The SK1D phase is very similar to the previously
studied motion of single skyrmions through periodic pinning lattices 
\cite{feilhauer_controlled_2020,reichhardt_quantized_2015,vizarim_skyrmion_2020}.
As the pinning density increases, the SK1D motion vanishes and
is replaced by soliton motion.
The gaps between the pinning centers diminish with increasing $\rho_p$,
making it impossible for the
interstitial skyrmion to move
unless it exchanges places with neighboring
skyrmions in a soliton-like fashion.

In the vacancy-containing sample
with $N_{sk}/N_p=0.99$, 
Fig.~\ref{Fig11}(b) shows that 
there is a monotonic decrease of the depinning threshold with increasing $\rho_p$.
As the sample density increases, the relative strength
of the skyrmion-skyrmion interactions increases compared to the pinning energy,
causing a suppression of the pinning threshold.
Soliton motion is completely lost for
$\rho_p<0.166$ when the large distance between adjacent pinning sites destroys
the collective behavior required to propagate a skyrmion through the sample.
For $\rho_p>0.166$, the extent of the soliton phase increases
with increasing pinning density, primarily due to the decrease in the depinning
threshold.
Both the
$N_{sk}/N_p=1.01$ and $N_{sk}/N_p=0.99$
samples show a similar
transition from 1D motion to 2D motion
since this transition is dominated
by the skyrmions in the strong pinning sites, which are the same in both systems.

\section{Guidance of soliton motion and skyrmion Hall angle reversal}

\begin{figure}[h]
    \centering
    \includegraphics[width=0.6\columnwidth]{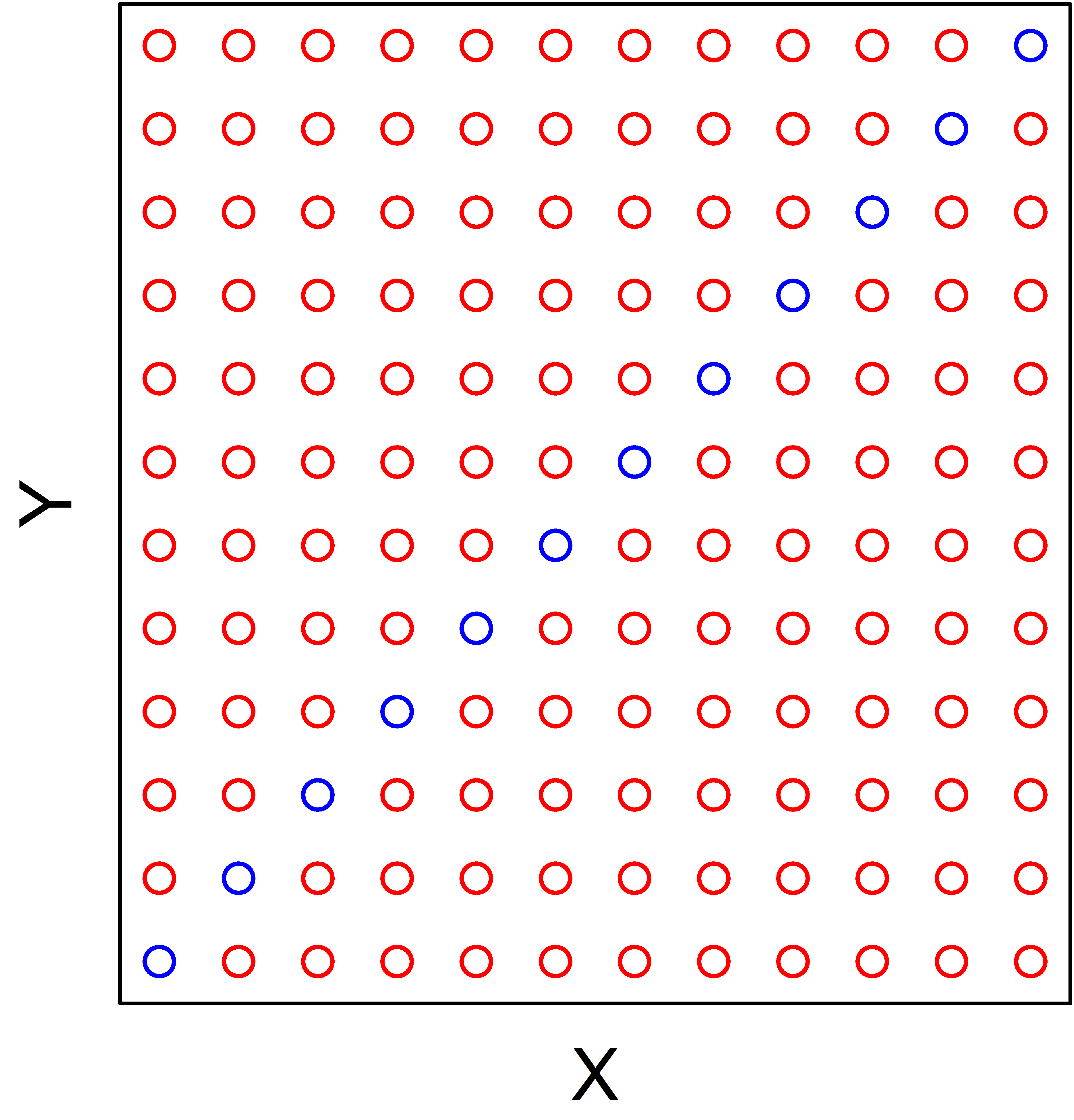}
    \caption{
Pinning site positions (red circles: strong pins; blue circles: weak pins)      
for a sample with a diagonal line of weak pinning oriented
at $+45^{\circ}$ with respect to the $x$ direction applied drive.
The weak pins have $U_p = 0.15$, the strong pins have
$U_p=1.0$,
and we set
    $\rho_p=0.373/\xi^2$.
}
    \label{Fig12}
\end{figure}

We have shown that soliton motion through skyrmion chains
can be enhanced depending on the choice of pinning density and
$\alpha_m/\alpha_d$.
We next ask whether
it is possible to guide the soliton motion along a specific direction.
When we introduced a line of weak pinning in the sample, the soliton
followed this line along
the $+x$ or $-x$ direction, depending on the value of $N_{sk}/N_p$,
even though this direction is not aligned with
the intrinsic Hall angle.
In other words,
guiding by the line of weak pinning potentials can overcome the skyrmion Hall angle.
To further explore this effect, we
change the skyrmion Hall angle
so that it is perpendicular to the guiding line of weaker pinning potentials.
As shown in Fig.~\ref{Fig12}, we place
the line of weak pinning centers along $\theta_{p}=+45^{\circ}$
with respect to the driving or $x$ direction.
By selecting $\alpha_m/\alpha_d = 1.0$,
we obtain an intrinsic skyrmion Hall angle of $\theta_{sk}^{\rm int}=-45^{\circ}$,
so that $\Delta \theta = \theta_{p} - \theta_{sk}^{\rm int} = 90^{\circ}$. 

\begin{figure}[h]
    \centering
    \includegraphics[width=1.0\columnwidth]{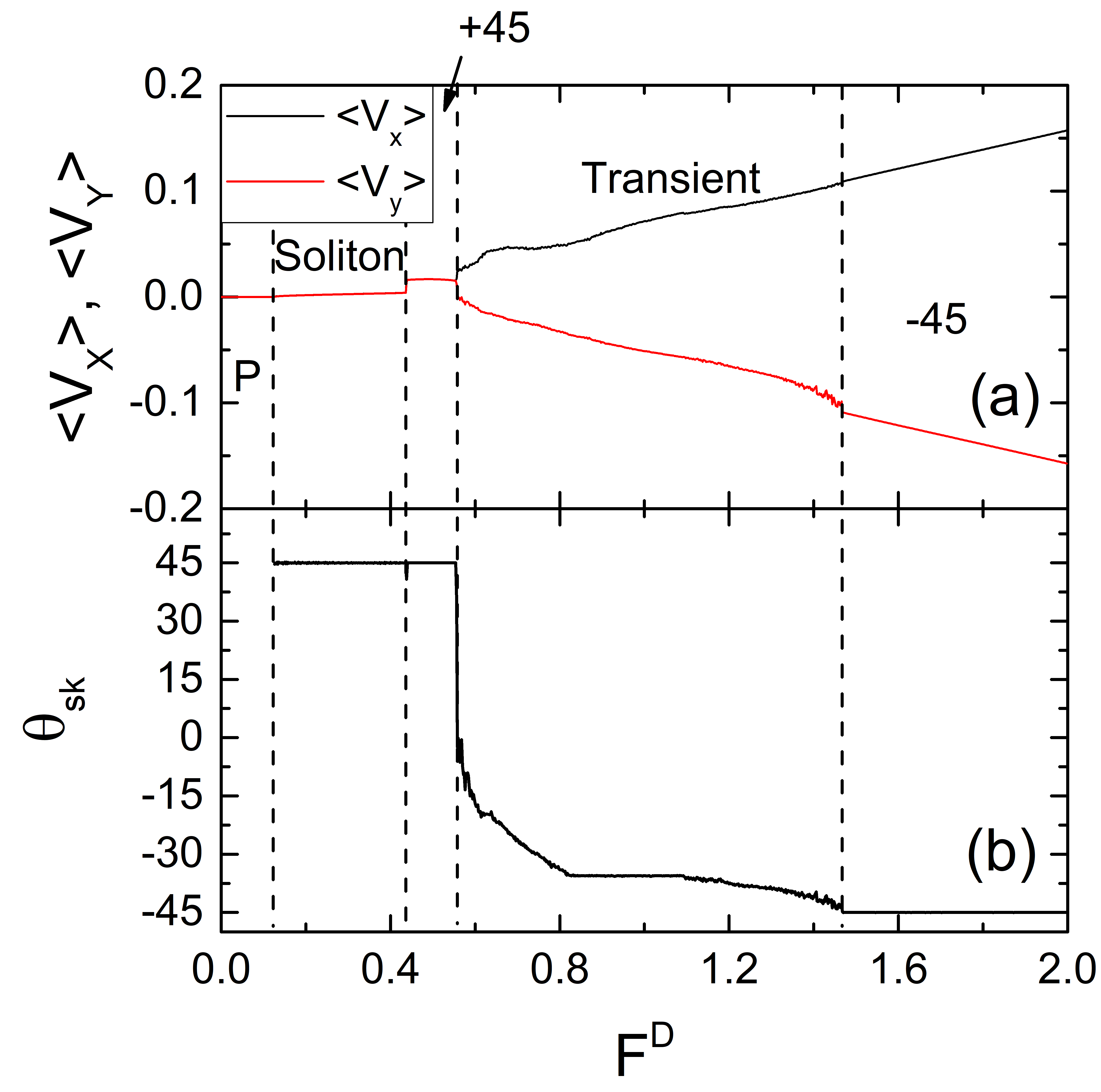}
\caption{
    (a) $\langle V_x\rangle$ (black) and $\langle V_y\rangle$ (red)
  versus
  $F^D$ for the
  sample illustrated in Fig.~\ref{Fig12}(b) with
  $N_{sk}/N_p = 1.01$, $\alpha_m/\alpha_d = 1.0$, and 
  $\rho_p = 0.373 \xi^2$.
  (b) The corresponding skyrmion Hall angle $\theta_{sk}$ versus $F^D$.
  $P$ indicates the pinned phase, Soliton is the soliton phase,
  $+45$ is the phase in which
  the skyrmions that have depinned from the weak pinning centers
  flow with $\theta_{sk}=+45^{\circ}$,
   Transient is the phase in which the skyrmion Hall angle is reversing, and
   $-45$ is the phase in which the motion is locked to $\theta_{sk}=-45^{\circ}$.
}
    \label{Fig13}
\end{figure}

\begin{figure}[h]
    \centering
    \includegraphics[width=1.0\columnwidth]{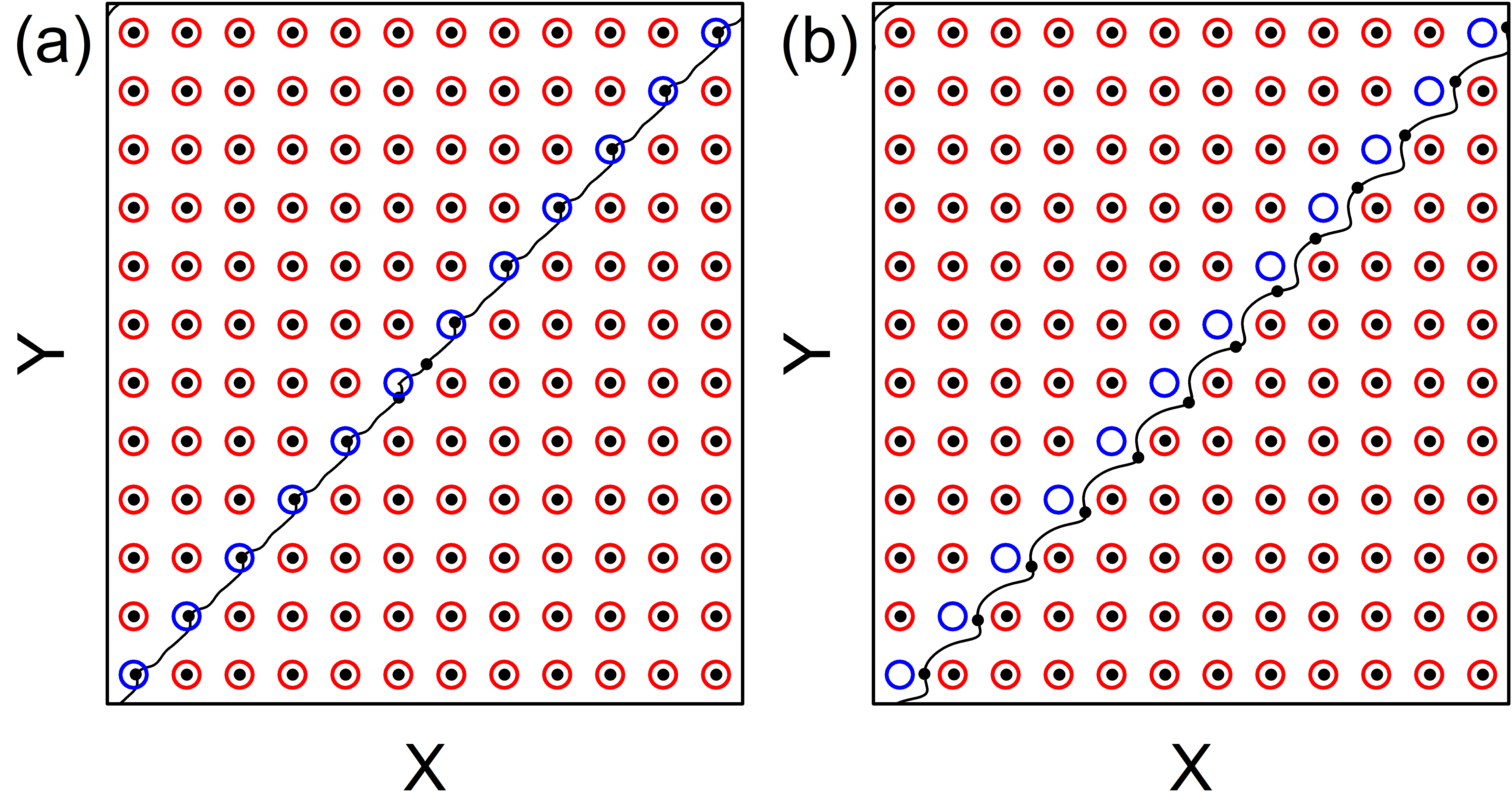}
    \includegraphics[width=1.0\columnwidth]{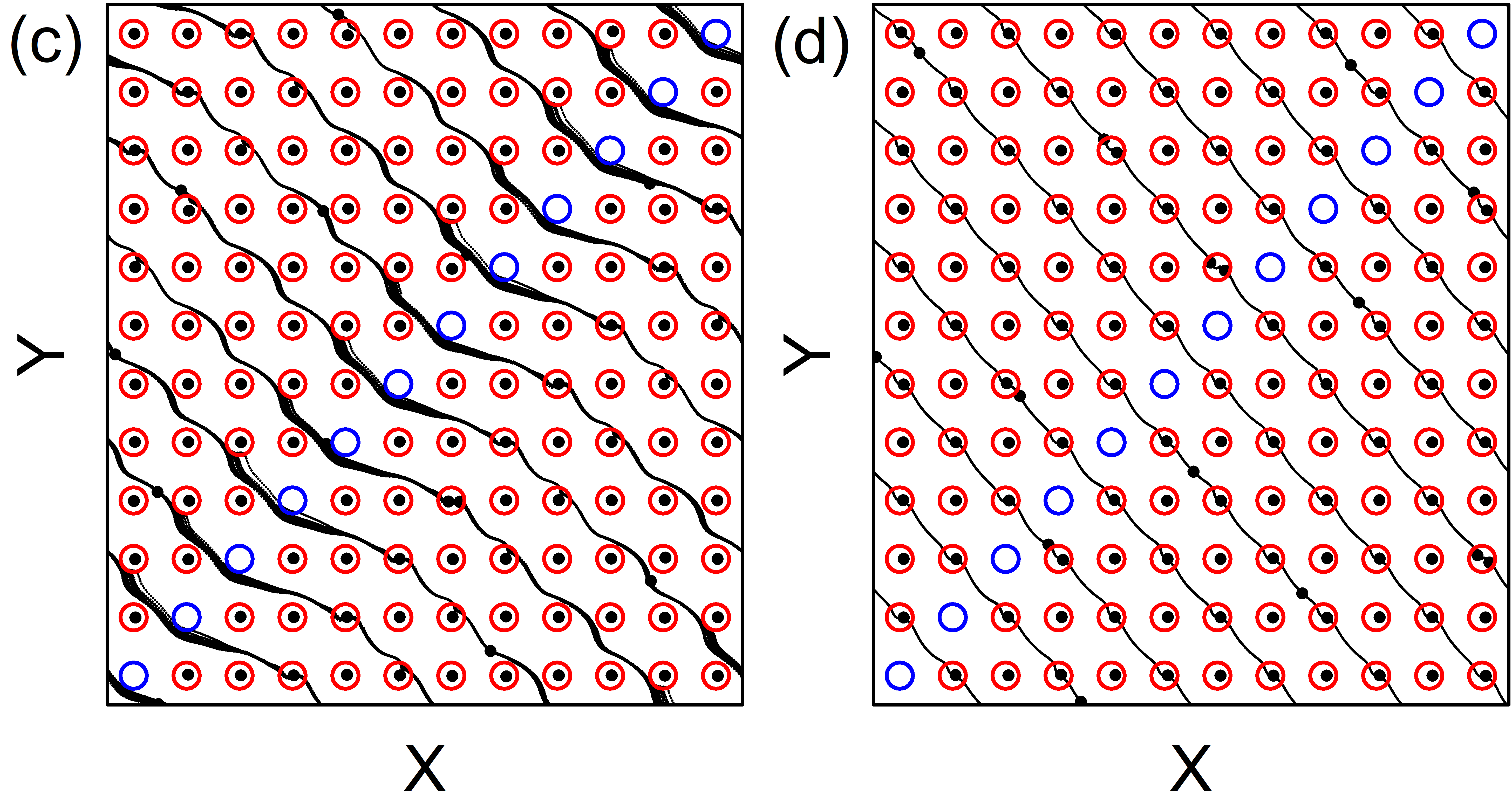}
\caption{
Pinning site positions (red circles: strong pins; blue circles: weak pins)
and the skyrmion trajectories (black lines)
for a sample with $N_{sk}/N_p = 1.01$,
$\alpha_m/\alpha_d = 1.0$, $\rho_p = 0.373\xi^2$,
weak pinning of $U_p = 0.15$, and  
strong pinning of $U_p=1.0$.
(a)
$F^D=0.25$, where a soliton flows along the line of weak pinning centers at
$\theta_{sk}=+45^{\circ}$.
(b)
At $F^D = 0.5$, the skyrmions trapped at the weak pinning sites depin and flow along
$\theta_{sk}=+45^{\circ}$.
(c)
$F^D = 1.0$, the transient phase, where the skyrmion Hall angle gradually reverses.
At this stage, the skyrmion Hall angle is $\theta_{sk}=-35.6^{\circ}$.
(d) 
At $F^D = 1.8$ the skyrmion Hall angle reversal is complete and
the skyrmions flow along $\theta_{sk}=-45^{\circ}$.
}
    \label{Fig14}
\end{figure}

For a sample with $N_{sk}/N_p = 1.01$,
we plot $\langle V_x \rangle$ and $\langle V_y \rangle$
versus $F^D$
in Fig.~\ref{Fig13}(a) and show the
corresponding
skyrmion Hall angle $\theta_{sk}^{\rm int}$ versus $F^D$ in
Fig.~\ref{Fig13}(b).
When $F^D < 0.12$, the skyrmions are in the pinned phase,
marked by the letter $P$ in Fig.~\ref{Fig13}(a).
As $F_D$ increases,
we observe a small and continuous increase of both
velocity components, which remain equal to each other so that
$\langle V_x \rangle = \langle V_y \rangle$.
Here the motion is occurring
at exactly $+45^{\circ}$ with respect to the driving direction and is following
the line of weak pinning centers.
An illustration of this soliton motion appears
in Fig.~\ref{Fig14}(a).
For $0.44<F^D<0.56$,
the skyrmion velocity components remain equal
to each other but do not change as the drive increases.
In this regime,
all of the skyrmions in the weak pinning sites
depin and flow with $\theta_{sk}=+45^{\circ}$, as shown in Fig.~\ref{Fig14}(b).
For $0.56<F^D<1.47$ we find a broad  transient phase
in which the skyrmion Hall angle slowly changes from $\theta_{sk}=+45^{\circ}$ to
$\theta_{sk}=-45^{\circ}$.
Here, the skyrmions in
the strong pinning sites remain pinned, but the depinned skyrmions
from the weak pinning sites begin to escape from the weak pinning channel
that is aligned with $\theta_{sk}=+45^\circ$ and instead start flowing along the
intrinsic skyrmion Hall angle of
$\theta_{sk}^{\rm int}=-45^{\circ}$.
A step in the skyrmion Hall angle at $\theta_{sk}=35.6^\circ$ appears around the
value
$F^D = 1.0$, corresponding to the flow state illustrated in
Fig.~\ref{Fig14}(c).
This motion is unstable and the magnitude of the skyrmion Hall angle continues
to increase once $F^D$ is raised above the step region.
The collective motion only becomes stable
once $F^D>1.46$, when
the skyrmions flow in an orderly
fashion along $\theta_{sk}=-45^{\circ}$, as shown in Fig.~\ref{Fig14}(d).
Here, some
of the skyrmions that were previously
trapped in the stronger pinning centers
have now depinned and serve to stabilize the flow.
The depinning of the remaining skyrmions occurs only
for drives higher than those considered here.

\section{Summary}

In this work we investigated the collective behavior of skyrmions at
zero temperature using a channel of weak pinning sites inserted into a periodic
lattice of strong pinning sites for
slightly incommensurate fillings.
We demonstrated that soliton motion can flow along the
chains of weak pinning sites.
The system displays
two types of soliton motion:
$(i)$ motion in the direction of drive for an interstitial soliton,
and $(ii)$ motion opposite to the direction of the drive for a
vacancy soliton.
These two types of soliton behave as if they have opposite charges,
and their direction of motion depends on their structure.
For a quasi-one dimensional sample,
both the soliton and the skyrmion motion are
strongly confined 
to the center axis of the sample
by the  repulsive barrier walls.
It is also possible to induce
soliton motion in 2D periodic lattices by providing a guiding channel in the form of
a line of weak pinning centers.
We show that the soliton motion is not strongly sensitive to the value of the
skyrmion Hall angle, but that it is
strongly affected by the pinning density.
At low pinning densities the skyrmions are too far
apart for collective behavior to appear and the soliton motion is destroyed.
As the pinning density increases, the skyrmion-skyrmion interactions become
relevant and a propagating soliton can be stabilized.
When we vary
$\alpha_m/\alpha_d$,
we find that the soliton motion is the most prominent if
the intrinsic skyrmion Hall angle is close to the
soliton direction of motion. Nevertheless, even for angular differences
as large as $90^\circ$,
the soliton motion persists over
a
range of applied drives, indicating that the soliton phase is robust.
In
a sample where the skyrmion Hall angle is perpendicular to
the weak pinning line,
the soliton motion is aligned with the weak pinning at $+45^\circ$.
As the external drive is increased, pinned skyrmions begin to depin and
the skyrmion Hall angle rotates from $+45^{\circ}$ to
$-45^{\circ}$ in order to align with the intrinsic Hall angle.
This indicates that at low drives the soliton motion can be guided,
while for higher drives the skyrmions follow the intrinsic Hall angle. 
Such behavior is of interest for technological applications where the skyrmion motion 
must be controlled precisely and must follow directions different than
the intrinsic skyrmion Hall angle.
The moving soliton could be used as an information carrier in logic devices rather
than the skyrmions themselves,
making it possible to transport information at drives much lower than those needed
to depin a chain of skyrmions. An advantage of this approach is that the solitons
do not exhibit a finite skyrmion Hall angle.

\acknowledgments
This work was supported by the US Department of Energy through
the Los Alamos National Laboratory and Research Foundation-Flanders (FWO).  
Los Alamos National Laboratory is
operated by Triad National Security, LLC, for the National Nuclear Security
Administration of the U. S. Department of Energy (Contract No. 892333218NCA000001).
N.P.V. acknowledges
funding from
Funda\c{c}\~{a}o de Amparo \`{a} Pesquisa do Estado de S\~{a}o Paulo - FAPESP (Grant 2017/20976-3).
J.C.B.S. acknowledges
funding from
Funda\c{c}\~{a}o de Amparo \`{a} Pesquisa do Estado de S\~{a}o Paulo - FAPESP (Grant 2021/04941-0).

\bibliography{mybib}
\end{document}